\def\year{2020}\relax
\documentclass[letterpaper]{article} 
\usepackage{aaai20}  
\usepackage{times}  
\usepackage{helvet} 
\usepackage{courier}  
\usepackage[hyphens]{url}  
\usepackage{graphicx} 

\usepackage{amsmath}
\usepackage{amssymb}
\usepackage{subfigure}
\usepackage{epstopdf}
\usepackage{algorithm}
\usepackage{algorithmic}
\usepackage{multirow}
\usepackage{array}
\usepackage[switch]{lineno}
\usepackage[titletoc,toc,page]{appendix}

\usepackage{graphicx}  
\title{Online Hashing with Efficient Updating of Binary Codes}
\author{\Large \textbf{Zhenyu Weng, Yuesheng Zhu}\\
Communication and Information Security Laboratory, Shenzhen Graduate School, Peking University\\
wzytumbler@pku.edu.cn, zhuys@pku.edu.cn
}
 
\begin{document}

\maketitle

\begin{abstract}
Online hashing methods are efficient in learning the hash functions from the streaming data. However, when the hash functions change, the binary codes for the database have to be recomputed to guarantee the retrieval accuracy. Recomputing the binary codes by accumulating the whole database brings a timeliness challenge to the online retrieval process. In this paper, we propose a novel online hashing framework to update the binary codes efficiently without accumulating the whole database. In our framework, the hash functions are fixed and the projection functions are introduced to learn online from the streaming data. Therefore, inefficient updating of the binary codes by accumulating the whole database can be transformed to efficient updating of the binary codes by projecting the binary codes into another binary space. The queries and the binary code database are projected asymmetrically to further improve the retrieval accuracy. The experiments on two multi-label image databases demonstrate the effectiveness and the efficiency of our method for multi-label image retrieval.
\end{abstract}

\section{Introduction}

With the explosive growth of the image data, finding the nearest neighbors to a query becomes a fundamental research problem in many computer vision applications~\cite{7915742}. Due to the efficiency in terms of space and time, hashing methods are widely used for nearest neighbor search, where the high-dimensional data are mapped to the binary codes~\cite{lu2017deep,wang2012semi,liu2017sequential}.

The hashing methods~\cite{Dizaji_2018_CVPR,7915742,he2019k} have achieved the promising performance. By learning the hash functions according to the data distribution and the label information, the hashing methods can map the high-dimensional data to the compact binary codes, and the similarity between the data can be preserved by the Hamming distance between the binary codes. However, these methods are batch-based hashing methods, which need to load the whole database into RAM to learn the hash functions. Due to the limited memory space, the deep hashing methods~\cite{DBLP:conf/aaai/LiDWXL19,DBLP:conf/aaai/SongHGXHS18} adopt minibatch-based stochastic optimization which requires multiple passes over a given database to learn the hash functions. However, they still have poor scalability for the large-scale application. Since the image data keep growing in the real application, to adapt to the change of the data, the hash functions need to be re-trained by accumulating the whole database when new data appear, which is time-consuming.

\begin{figure*}[t]
\centering
\begin{tabular}{cc}
 \includegraphics[width=0.9\columnwidth]{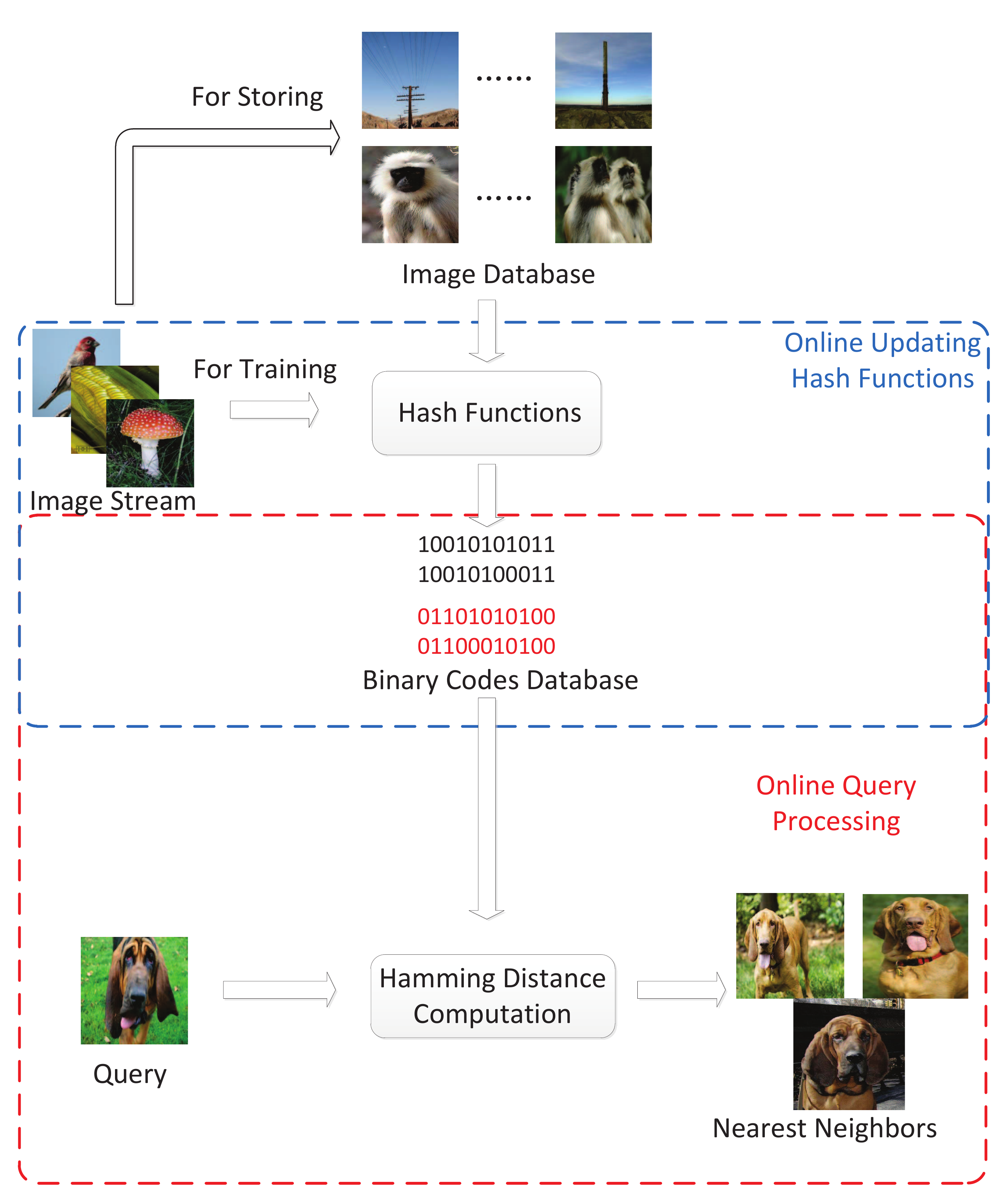}&
  \includegraphics[width=0.9\columnwidth]{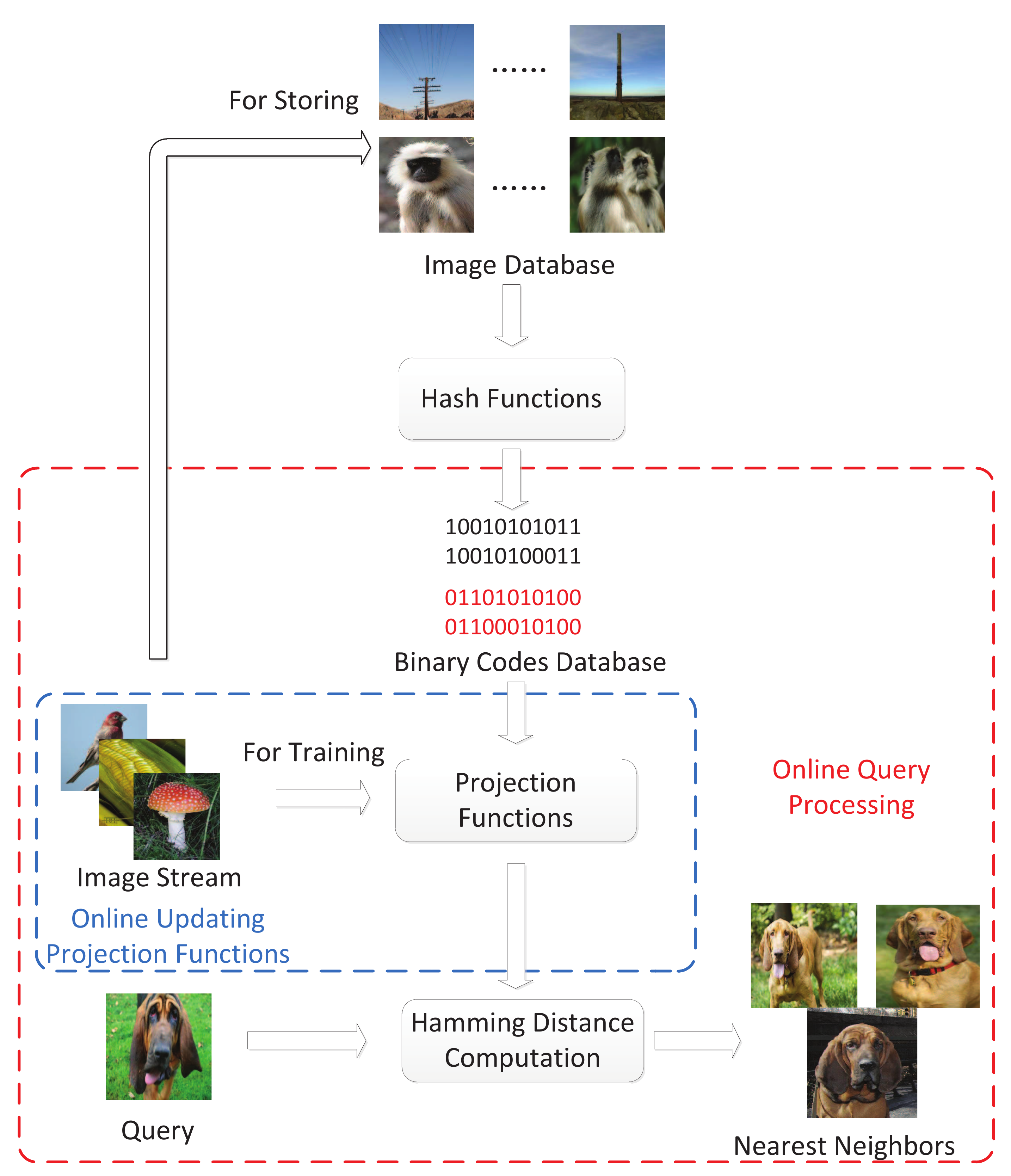}\\
{\small (a) Updating hash functions}  &  {\small (b) Updating projection functions}
\end{tabular}
\caption{Comparison between the online hashing framework that updates hash functions and the proposed online hashing framework that updates projection functions. The blue box shows the online function updating and the red box shows the online query processing. (a) The online hashing framework updates the hash functions from the streaming data. (b) The proposed framework fixes the hash functions and updates the projection functions from the streaming data.}
\label{fig:res0}
\end{figure*}

Online hashing methods~\cite{cakir2015adaptive,cakir2017online,Cakir2017MIHash,7907165,Lin:2018:SOH:3240508.3240519,DBLP:conf/aaai/LinJLSWW19,DBLP:conf/aaai/XieSZ16,ijcai2017-437} are recently proposed to employ online learning techniques on the streaming data. By learning the hash functions from the streaming data, online hashing methods can adapt to the data variations with low computational complexity. According to their application scenarios, the online hashing methods can be categorized as single image modality retrieval~\cite{cakir2017online,Cakir2017MIHash,DBLP:conf/aaai/LinJLSWW19} and cross-modal retrieval~\cite{DBLP:conf/aaai/XieSZ16}. In this paper, we focus on the single image modality retrieval.

Online Sketch Hashing (OSH)~\cite{leng2015online} maintains a data sketch to preserve the properties of the streaming data and learns the hash functions from the data sketch. But it is an online unsupervised hashing method which cannot use the label information to improve the retrieval accuracy. To take advantage of the label information, following the Passive-Aggressive algorithm~\cite{crammer2006online}, Online Kernel-based Hashing (OKH)~\cite{7907165} defines a structured similarity loss function and updates the hash functions according to the loss function when receiving a new pair of the data and their similarity label. Mutual Information Hashing (MIH)~\cite{Cakir2017MIHash} uses the mutual information as the objective function and updates the hash functions according to the objective function with each incoming data point and its corresponding class label. Online Supervised Hashing (OSupH)~\cite{cakir2017online} generates error correcting output codes according to the label information and uses these codes to guide the learning process of the hash functions from the streaming data. Balanced Similarity for Online Discrete Hashing
(BSODH)~\cite{DBLP:conf/aaai/LinJLSWW19} preserves the correlation between the streaming data and
the existing database via an asymmetric graph regularization and updates the hash functions with this correlation.

Although some progress has been made, the online hashing methods discussed above simply focus on modeling the hash functions to be updated efficiently. They ignore the problem that the binary codes for the indexed data have to be updated to guarantee the retrieval accuracy whenever the hash functions change, and updating the binary codes frequently by accumulating the ever-increasing database inevitably blocks the timeliness of the online retrieval process. Though the previous work, $e.g.$, MIH~\cite{Cakir2017MIHash} tries to solve this problem by introducing a trigger update module to reduce the frequency of recomputing the binary codes. However, the calculation in the trigger update module is time-consuming. Cross-Modal Hashing (OCMH)~\cite{DBLP:conf/aaai/XieSZ16} also detects the inefficiency of accumulating the whole database to update the binary codes and improves the updating efficiency by representing the binary codes with the permanent latent sharing codes and the dynamic transfer matrix. But this method is designed for the cross-modal retrieval application, and cannot be directly used for single image modality retrieval.

In this paper, we propose a novel online hashing framework to update the binary codes without accumulating the whole database. Different from the above online hashing methods that learn the hash functions from the streaming data, in our framework, the hash functions are fixed and the projection functions are introduced to learn from the streaming data. Hence, inefficient updating of the binary codes by accumulating the whole database can be transformed to efficient updating of the binary codes by projecting the binary codes into another binary space. The comparison of our framework against the current online hashing methods is shown in Fig.~\ref{fig:res0}. For the current online hashing methods, they update the hash functions from the streaming data. In contrast, in our framework, the hash functions are fixed, and the projection functions are updated from the streaming data.

To sum up, our contribution in this paper is three fold:
\begin{itemize}
  \item By introducing the projection functions, a novel online hashing framework is proposed to update the binary codes without accumulating the whole database.
  \item Different from most online hashing methods using the single-label information to learn the hash functions, our framework takes advantage of the multi-label information to learn the projection functions.
  \item To further improve the retrieval accuracy, the projection functions for the queries and the binary code database are learnt in an asymmetric way.
\end{itemize}

\section{Online Hashing with Efficient Updating}
\subsection{Hash Function}
Given a set of image features $X = \{ {\bold{x}_1},...,{\bold{x}_N}\} $, where ${\bold{x}_i} \in {\mathbb R}^D$ and $D$ is the dimensionality of the feature vector, hashing methods aim to map $\bold{x}_i$ to a $K$-bit binary code $\bold{h}_i \in {\{  - 1,1\} ^K}$. For a pair of data points $\bold{x}_i$ and $\bold{x}_j$, the hamming distance between them is calculated as
\begin{equation}
{S_H}({\bold{x}_i},{\bold{x}_j}) = {\bold{h}_i} \otimes {\bold{h}_j},
\end{equation}
where $\otimes$ is an xor operation.

Following ~\cite{Cakir2017MIHash,DBLP:conf/aaai/LinJLSWW19}, we adopt the linear projection-based hash functions. To encode a given data point into a $K$-bit binary code, the hash function is defined as
\begin{equation}
\bold{h}_i = {\mathop{\rm sgn} (\bold{W}^T\bold{x}_i + \bold{b})},
\end{equation}
where $\bold{W} \in {{\mathbb R}^{D \times K}}$ is a projection matrix and $\bold{b} \in {\mathbb R}^K$ is a threshold vector.

Different from other online hashing methods that update the projection matrix $\bold{W}$, in our method, we fix $\bold{W}$ and learn the projection functions to map the generated binary code ${\bold{h}_i}$ into another binary space, which will be elaborated in the next subsection.

We use Iterative Quantization (PCA-ITQ)~\cite{gong2013iterative} to learn $\bold{W}$. PCA-ITQ is an unsupervised hashing method which can preserve the similarity between data effectively for both the seen and the unseen data. By setting the $k^{th}$ threshold value $b_k$ of $\bold{b}$ as the negative mean of the data on the $k^{th}$ dimension and learning a projection matrix $\bold{W}$, PCA-ITQ can minimize the quantization error between the projected data and the corresponding binary codes. PCA-ITQ is traditionally a batch-based hashing method. Nevertheless, recent advance ~\cite{7936671} shows that this hashing method can preserve the data similarity well by just using a few data points as the training data for learning the hash functions. Hence, we adopt PCA-ITQ to learn the hash functions in the initial stage. Following~\cite{7907165}, we assume that at least $m$ data points have been provided in the initial stage, and the hash functions are learnt according to these data points.

\subsection{Projection Function}
Assume a stochastic environment in which data $\bold{x}$ come sequentially along with the class label $y \in \Upsilon $, where  $\Upsilon $ is the label space. For a pair of data points $\bold{x}_i$ and $\bold{x}_j$, their corresponding binary codes are $\bold{h}_i$ and $\bold{h}_j$, respectively. In our method, as the hash functions are fixed, to use the label information, each binary code $\bold{h}$ is mapped to another binary code $\bold{g} \in {\{- 1,1\} ^{K}}$, which is defined as
\begin{equation}
\bold{g} = {\mathop{\rm sgn}} ({\bold{P}^T}\bold{h}),
\end{equation}
where $\bold{P} \in {\mathbb R}^{K \times K}$ is a projection matrix.

Hence, the similarity between the data in our method is calculated as
\begin{equation}
S({\bold{x}_i},{\bold{x}_j}) = {\mathop{\rm sgn}} ({\bold{P}^T}{\bold{h}_i}) \otimes {\mathop{\rm sgn}} ({\bold{P}^T}{\bold{h}_j}) = {\bold{g}_i} \otimes {\bold{g}_j}.
\end{equation}

Inspired by OKH~\cite{7907165} and OSupH~\cite{cakir2017online} which generate the ideal binary code according to the supervised information to guide the updating of the hash functions, we generate the ideal binary code according to the class label $y$ to guide the updating of the projection functions.

Different from OKH and OSupH that consider about the single-label situation, our method considers about the multi-label situation, and generates the ideal binary code that contains the multi-label information. In the label space of $C$ classes, we use one-hot coding, and thus $\Upsilon  = \{ \bold{y}|\bold{y} \in {\{ 0,1\} ^C}\} $ where the $i^{th}$ entry is 1 if the data point belongs to the $i^{th}$ class.  It has been proved in~\cite{7636996} that Locality Sensitive Hashing (LSH)~\cite{andoni2006near} can preserve the semantic similarity among label-based representation. If two labels are similar, the collision probability of the corresponding binary codes is greater than 0.5. Otherwise, the collision probability  is less than 0.5. Hence, for a given data point $\bold{x}_i$ and its label $\bold{y}_i$, the ideal binary code $\bold{g}_i^*$ is
\begin{equation}
\bold{g}_i^* = {\mathop{\rm sgn}} ({\bold{L}^T}{\bold{y}_i}),
\end{equation}
where $\bold{L} \in {{\mathbb R}^{C \times K}}$ is a random matrix of which each column is sampled from the normal Gaussian distribution ${\cal N} (0, \bold{I})$.

The projection matrix $\bold{P} \in {{\mathbb R}^{K \times K}}$ is composed of projection vectors $[\bold{p}_1,...,\bold{p}_{K}]$ and each vector corresponds to one bit of the ideal binary code. Different from other online hashing methods~\cite{cakir2015adaptive,Cakir2017MIHash} that treat the bits collectively, we treat the bits independently. The value of each bit is 1 or -1, and the projection vectors are learnt independently.

Inspired by~\cite{crammer2006online}, we update the projection vectors from the streaming data in a Passive-Aggressive (PA) way. For notational simplicity, we use $\bold{p}$ to denote the $k^{th}$ vector $\bold{p}_k$ as each projection vector is learnt independently, and $g_i^*$ to denote the corresponding bit of the ideal binary code $\bold{g}^*$ for the label $\bold{y}_i$ of the $i^{th}$ data point $\bold{x}_i$.  The loss function for each coming data point $\bold{x}_i$ is defined as
\begin{equation}
l({g_i^*},{{\bf{h}}_i}) = \left\{ \begin{array}{l}
0\quad \quad \quad \quad \quad {g_i^*}({{\bf{p}}^T}{{\bf{h}}_i}) \ge 1,\\
1 - {g_i^*}({{\bf{p}}^T}{{\bf{h}}_i})\quad otherwise,
\end{array} \right.
\end{equation}
where $\bold{h}_i$ is the corresponding binary code of $\bold{x}_i$.

For brevity, we use $l_i$ to denote the loss $l({g_i^*},{{\bf{h}}_i})$ for the $i^{th}$ data point. At first, $\bold{p}$ is initialized with a random value. Then, for each round $i$, we solve the following convex problem with soft margin:
\begin{equation}
\begin{array}{l}
{{\bf{p}}_{i}} = \arg {\kern 1pt} \;\min \;{\textstyle{1 \over 2}}||{\bf{p}} - {{\bf{p}}_{i-1}}||^2 + C\xi \\
\;\;\;\;\;s.t.\quad {l_i} \le \xi \;and\;\xi  \ge 0,
\end{array}
\end{equation}
where $||\cdot||$ is the Euclidean norm and $C$ is the aggressiveness parameter that controls the trade-off between maintaining $\bold{p}$ close to the previous parameter $\bold{p}_{i-1}$ and minimizing the loss on the current loss $l_i$.

When $l_i$ = 0, $\bold{p}_{i}$ = $\bold{p}_{i-1}$ satisfies Eqn.(7) directly. Otherwise, we define the Lagrangian as:
\begin{equation}
\begin{split}
L(\bf{p},\tau,\lambda,\xi) &= {\textstyle{1 \over 2}}||{\bf{p}} - {{\bf{p}}_{i - 1}}||^2 + C\xi \\
                           &+ \tau (1-\xi-{g_i^*}({{\bf{p}}^T}{{\bf{h}}_i})) - \lambda\xi,\\
\end{split}
\end{equation}
with $\tau \ge 0$ and  $\lambda \ge 0$ are Lagrange multipliers.

Let ${\textstyle{\partial L(\bf{p},\tau,\lambda,\xi)  \over \partial \bold{p}}} = 0$. We have
\begin{equation}
{\textstyle{\partial L(\bf{p},\tau,\lambda,\xi)  \over \partial \bold{p}}} = \bold{p} - \bold{p}_{i-1} - \tau{g^*_i}{\bf{h}}_i = 0.
\end{equation}

Let ${\textstyle{\partial L(\bf{p},\tau,\lambda,\xi)  \over \partial \xi}} = 0$. We have
\begin{equation}
{\textstyle{\partial L(\bf{p},\tau,\lambda,\xi)  \over \partial \xi}} = C - \tau - \lambda = 0.
\end{equation}

As $\lambda \ge 0$, $\tau \le C$.

Plugging Eqn.(9) and (10) back in Eqn.(8), we obtain
\begin{equation}
L(\tau)= -{\textstyle{1 \over 2}}\tau^2(g^*_i)^2||\bold{h}_i||^2+\tau(1-{g_i^*}({{\bf{p}}_{i-1}^T}{{\bf{h}}_i})).
\end{equation}

As $g^*_i \in \{1, -1\}$, $(g^*_i)^2=1$. As $\bold{h}_i \in \{-1,1\}^K$, $||\bold{h}_i||^2 = K$. Let ${\textstyle{\partial L(\tau)  \over \partial \tau}} = 0$. We have
\begin{equation}
\tau = {\textstyle{ 1-{g_i^*}({{\bf{p}}_{i-1}^T}{{\bf{h}}_i})  \over ||\bold{h}_i||^2}} = {\textstyle{ 1-{g_i^*}({{\bf{p}}_{i-1}^T}{{\bf{h}}_i})  \over K}}.
\end{equation}

Hence, the optimal $\bold{p}$ is
\begin{equation}
{{\bf{p}}_i} = {{\bf{p}}_{i - 1}} + \tau {g^*_i}{{\bf{h}}_i},
\end{equation}
where
\begin{equation}
\tau  = \min \{ C,\frac{{1-{g_i^*}({{\bf{p}}_{i-1}^T}{{\bf{h}}_i})}}{{K}}\}.
\end{equation}

Hence, when given a query $\bold{q}$, the similarity between the query and the binary code is calculated as
\begin{equation}
S(\bold{q},{\bold{x}_i}) = {\mathop{\rm sgn}} ({\bold{P}^T}{\bold{h}_q}) \otimes {\mathop{\rm sgn}} ({\bold{P}^T}{\bold{h}_i}) = {\bold{g}_q} \otimes {\bold{g}_i},
\end{equation}
where $\bold{h}_q$ is the corresponding binary code of the query $\bold{q}$.

\subsection{Loss Bound}
Based on the Lemma 1 in ~\cite{crammer2006online}, we can obtain the following theorem.

$\bold{Theorem}$ $\bold{1}$ Let ($\bold{h}_1$, $g_1^*$),$\cdots$,($\bold{h}_i$, $g_i^*$) be a sequence of pair-wise examples with label $g_i^* \in \{-1,1\}$ for all $i$. As before, Eqn.(6) denotes by $l_i=l(\bold{p}; (\bold{h}_i$, $g_i^*))$ the instantaneous loss suffered by our method on round $i$. Let $\bold{u}$ be an arbitrary vector that $\bold{u} \in {\mathbb R}^{K}$, and define $l_i^*=l(\bold{u}; (\bold{h}_i$, $g_i^*))$ as the loss suffered by $\bold{u}$. Then, for any vector $\bold{u} \in {\mathbb R}^K$, the number of prediction mistakes made by our method on this sequence of examples is bounded by
\begin{equation}
\max \{ {K},1/{{C}}\} (||\bold{u}|{|^2} + 2C\sum\limits_{i = 1}^t {l_i^*} ).
\end{equation}

$\bold{Proof}$ According to Eqn.(6), if our method makes a prediction mistake on round $i$ then $l_i \ge 1$. Therefore, according to Eqn.(14), we have
\begin{equation}
\min \{ 1/{K},{\rm{C}}\}  \le {\tau _i}{l_i}.
\end{equation}

Let $M$ be the number of prediction mistakes made on the entire sequence. We have,
\begin{equation}
\min \{ 1/{K},{\rm{C}}\} M \le \sum\limits_{i = 1}^t {{\tau _i}{l_i}}.
\end{equation}

According to Eqn.(12), we know that
\begin{equation}
\tau_i||\bold{h}_i|{|^2} \le {l_i},
\end{equation}
and
\begin{equation}
\tau_il_i^* \le Cl_i^*.
\end{equation}

According to $\bold{Lemma}$ 1 in ~\cite{crammer2006online}, we have
\begin{equation}
\sum\limits_{i = 1}^t {{\tau _i}(2{l_i} - {\tau _i}||{\bold{h}_i}|{|^2} - 2l_i^*) \le ||\bold{u}|{|^2}}.
\end{equation}

Plugging Eqn.(19) and (20) into Eqn.(21), we obtain
\begin{equation}
\sum\limits_{{\rm{i}} = 1}^t {{\tau _i}{l_i}}  \le ||\bold{u}|{|^2} + 2C\sum\limits_{i = 1}^t {l_i^*}.
\end{equation}

Combining Eqn.(18) with Eqn.(22), we obtain

\begin{equation}
\min \{ 1/{K},C\} M \le ||\bold{u}|{|^2} + 2C\sum\limits_{i = 1}^t {l_i^*}.
\end{equation}

Multiplying both sides of the above by $\max \{ {K},1/{{C}}\}$, we have
\begin{equation}
M \le \max \{ {K},1/{{C}}\} (||\bold{u}|{|^2} + 2C\sum\limits_{i = 1}^t {l_i^*} ).
\end{equation}

\subsection{Asymmetric Projection}

In~\cite{NIPS2013_5017}, it shows that treating the query points and the database points asymmetrically can improve the search accuracy of the binary codes. Hence, we treat the query points and the database points in an asymmetric way by learning different projection functions. Instead of using the projection matrix $\bold{P}$ after hashing the query into a binary code, we learn a projection matrix $\bold{R}\in {\mathbb R}^{D \times K}$ that can be directly applied on the original query feature $\bold{q}\in {\mathbb R}^D$. Hence, the similarity between the query and the binary code is calculated as
\begin{equation}
S(\bold{q},{\bold{x}_i}) = {\mathop{\rm sgn}} ({\bold{R}^T}\bold{q}) \otimes {\mathop{\rm sgn}} ({\bold{P}^T}{\bold{h}_i}) = {\bold{g}_q} \otimes {\bold{g}_i},
\end{equation}
where $\bold{h}_i$ is the corresponding binary code of $\bold{x}_i$.

The projection matrix $\bold{R}$ is also composed of projection vectors $[\bold{r}_1,...,\bold{r}_{K}]$ and each vector corresponds to one bit of the ideal binary code. For notational simplicity, we use $\bold{r}$ to denote the $k^{th}$ vector $\bold{r}_k$. The loss function of $\bold{r}$ for each coming data point $\bold{x}_i$ is defined as:
\begin{equation}
\hat l({g_i^*},{{\bf{x}}_i}) = \left\{ \begin{array}{l}
0\quad \quad \quad \quad \quad {g_i^*}({{\bf{r}}^T}{{\bf{x}}_i}) \ge 1,\\
1 - {g_i^*}({{\bf{r}}^T}{{\bf{x}}_i})\quad otherwise.
\end{array} \right.
\end{equation}

For brevity, we use $\hat l_i$ to denote the loss $\hat l({g_i^*},{{\bf{x}}_i})$ for the $i^{th}$ data point. Then, for each round $i$, we solve the following convex problem with soft margin:
\begin{equation}
\begin{array}{l}
{{\bf{r}}_i} = \arg {\kern 1pt} \;\min \;{\textstyle{1 \over 2}}||{\bf{r}} - {{\bf{r}}_{i - 1}}||^2 + C\xi \\
\;\;\;\;\;s.t.\quad {\hat l_i} \le \xi \;and\;\xi  \ge 0.
\end{array}
\end{equation}

Therefore, the optimal $\bold{r}$ is
\begin{equation}
{{\bf{r}}_i} = {{\bf{r}}_{i - 1}} + \tau g_i^*{{\bf{x}}_i},
\end{equation}
where
\begin{equation}
\tau  = \min \{ C,\frac{{1 - {g_i^*}({{\bf{r}}_{i-1}^T}{{\bf{x}}_i})}}{{||{{\bf{x}}_i}|{|^2}}}\}.
\end{equation}

The derivation of Eqn.(28) is similar to the derivation of Eqn.(13), and is provided in Appendix A of the supplementary file.

By assuming $||{\bf{x}}_i|| \le R$, the prediction mistake of our method on the queries is bounded according to $\bold{Theorem}$ $\bold{2}$.

$\bold{Theorem}$ $\bold{2}$ Let ($\bold{x}_1$, $g_1^*$),$\cdots$,($\bold{x}_i$, $g_i^*$) be a sequence of pair-wise examples with label $g_i^* \in \{-1,1\}$ for all $i$. As before, Eqn.(26) in the manuscript denotes by $\hat l_i=\hat l(\bold{r}; (\bold{x}_i$, $g_i^*))$ the instantaneous loss suffered by our algorithm on round $i$. Let $\bold{u}$ be an arbitrary vector that $\bold{u} \in {\mathbb R}^{K}$, and define $\hat l_i^*=\hat l(\bold{u}; (\bold{x}_i$, $g_i^*))$ as the loss suffered by $\bold{u}$. Then, by assuming $||{\bf{x}}_i|| \le R$, for any vector $\bold{u} \in {\mathbb R}^K$, the number of prediction mistakes made by our method on this sequence of examples is bounded by
\begin{equation}
\max \{ {R^2},1/C\} (||\bold{u}|{|^2} + 2C\sum\limits_{i = 1}^t {\hat l_i^*} ).
\end{equation}

$\bold{Proof}$  The proof of $\bold{Theorem}$ $\bold{2}$ is provided in Appendix B of the supplementary file.

As our method can update the binary codes efficiently, our method is called Online Hashing with Efficient Updating (OHWEU). The pseudo-code of our method is shown in Algorithm 1.

\begin{algorithm}
\caption{OHWEU}               
\label{alg1}                      
\begin{algorithmic}[1]
\REQUIRE streaming data $(\bold{x}_i,\bold{y}_i)$, $C$, $K$               
\ENSURE $\bold{P}$ and $\bold{R}$  
\STATE Learn $\{ {\bold{w}_k,b_k}\} _{k = 1}^K$ by PCA-ITQ in the initial stage
\STATE Initialize $\bold{P}^0=[\bold{p}^0_1,...,\bold{p}^0_{K}]$ and $\bold{R}^0=[\bold{r}^0_1,...,\bold{r}^0_{K}]$
\FOR {$i=1,2,...$}
\STATE Obtain the binary code $\bold{g}^*_i$ according to Eqn.(5)
\FOR {$j=1$ to $K$}
\STATE Update $\bold{p}^i_j$ according to Eqn. (13)
\STATE Update $\bold{r}^i_j$ according to Eqn. (28)
\ENDFOR
\ENDFOR

\end{algorithmic}
\end{algorithm}

\section{Experiments}
\subsection{Datasets}
We compare our method on two commomly-used multi-label image datasets, MS-COCO and NUS-WIDE.

(a).MS-COCO dataset~\cite{10.1007/978-3-319-10602-1_48}. The MS-COCO dataset is a multi-label dataset consisting of training images and 40,504 validation images. Each image is labeled by some of the 80 concepts. We obtain 122,218 images after we filter the images that do not contain any concept label and combine the training images with the validation images. Each images is represented by a 4096-D feature extracted from the fc7 layer of a VGG-16 network~\cite{Simonyan14c} pretrained on ImageNet~\cite{imagenet_cvpr09}. We randomly take 4000 images as the queries, and the rest for training and searching.

(b). NUS-WIDE dataset~\cite{Chua:2009:NRW:1646396.1646452}. The NUS-WIDE dataset is also a multi-label dataset, which consists of 269,648 web images associated with tags. Following~\cite{Kang:2016:CSB:3015812.3015994}, we only select the images that belong to the 21 most frequent concepts. Hence, we have 195,834 images. Each images is represented by a 4096-D feature extracted from the fc7 layer of a VGG-16 network~\cite{Simonyan14c} pretrained on ImageNet. We randomly take 2000 images as the queries, and the rest for training and searching.

The experiments are run on a computer with CPU I7, 24-GB memory. Mean Average Precision (mAP) is used to measure the performance of the online hashing methods. If two images share at least one common label, they are defined as a groundtruth neighbor. The results are averaged by repeating the experiments 3 times.

\begin{figure}[t]
\centering
\begin{tabular}{cc}
 \includegraphics[width=0.455\columnwidth]{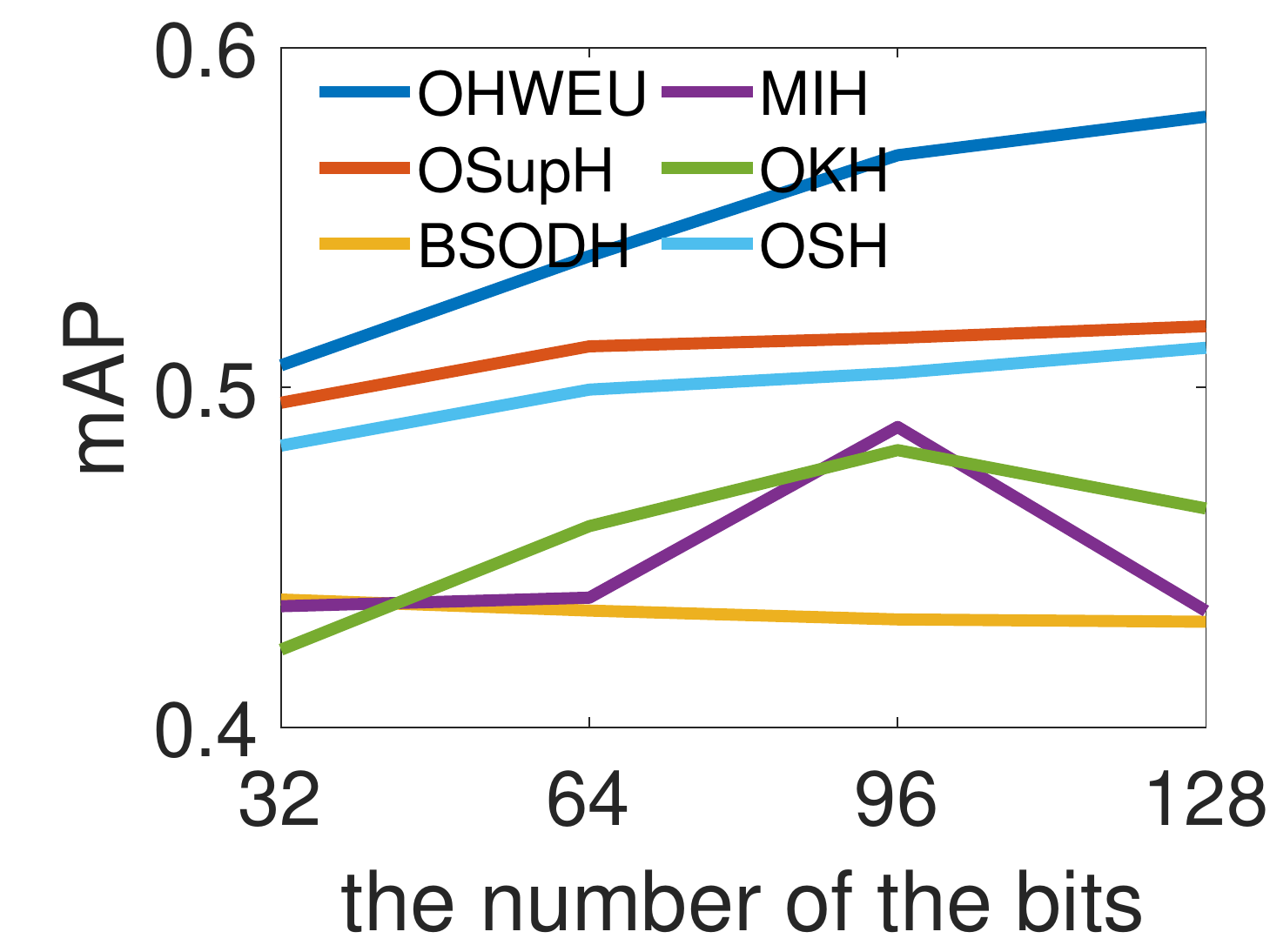}&
 \includegraphics[width=0.455\columnwidth]{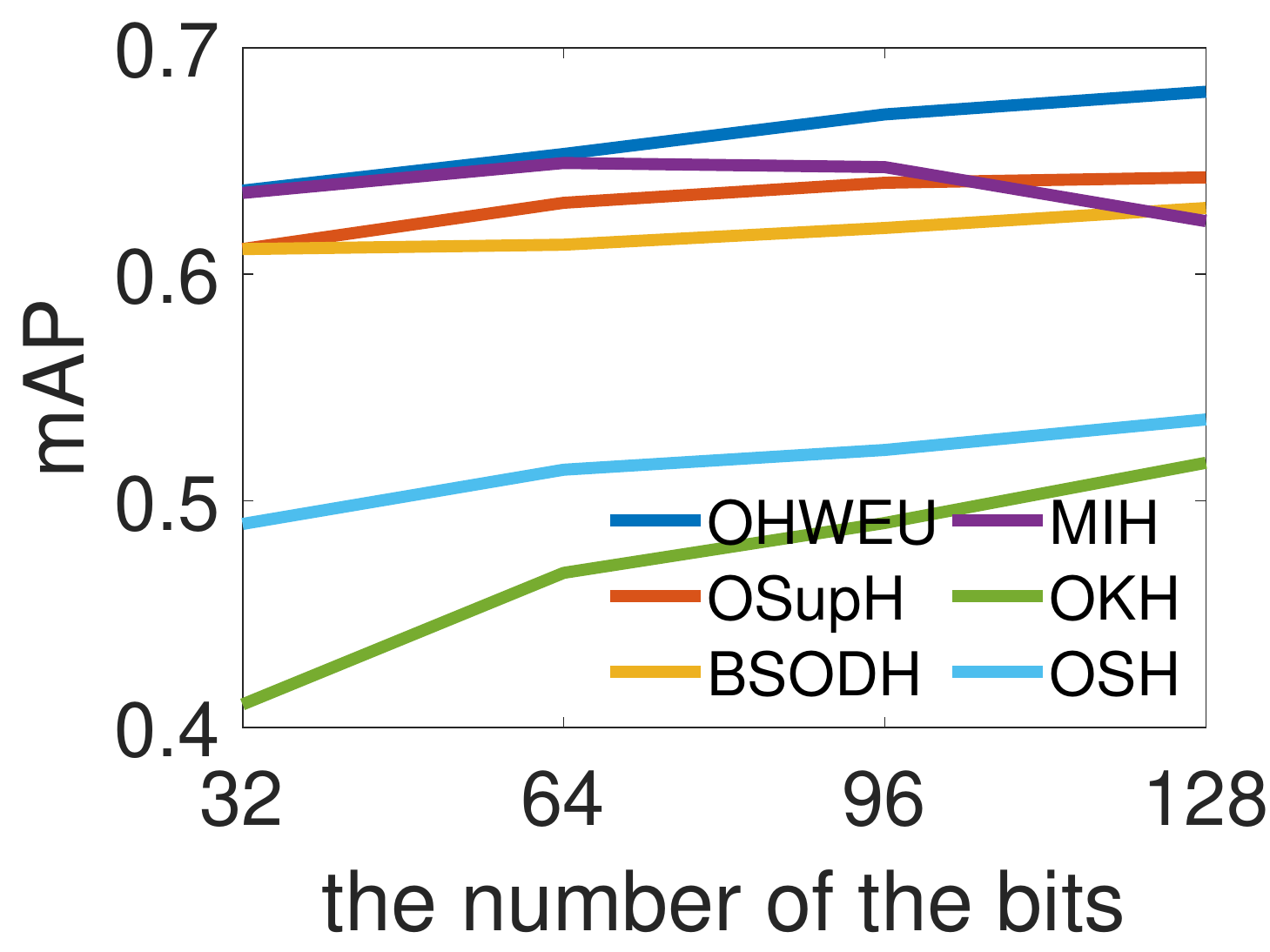}\\
{\small (a) MS-COCO}  &  {\small (b) NUS-WIDE}
\end{tabular}

\caption{The mAP results of different hashing methods.}
\label{fig:res2}

\end{figure}

\begin{figure*}[t]
\centering
\begin{tabular}{cccc}
 \includegraphics[width=0.48\columnwidth]{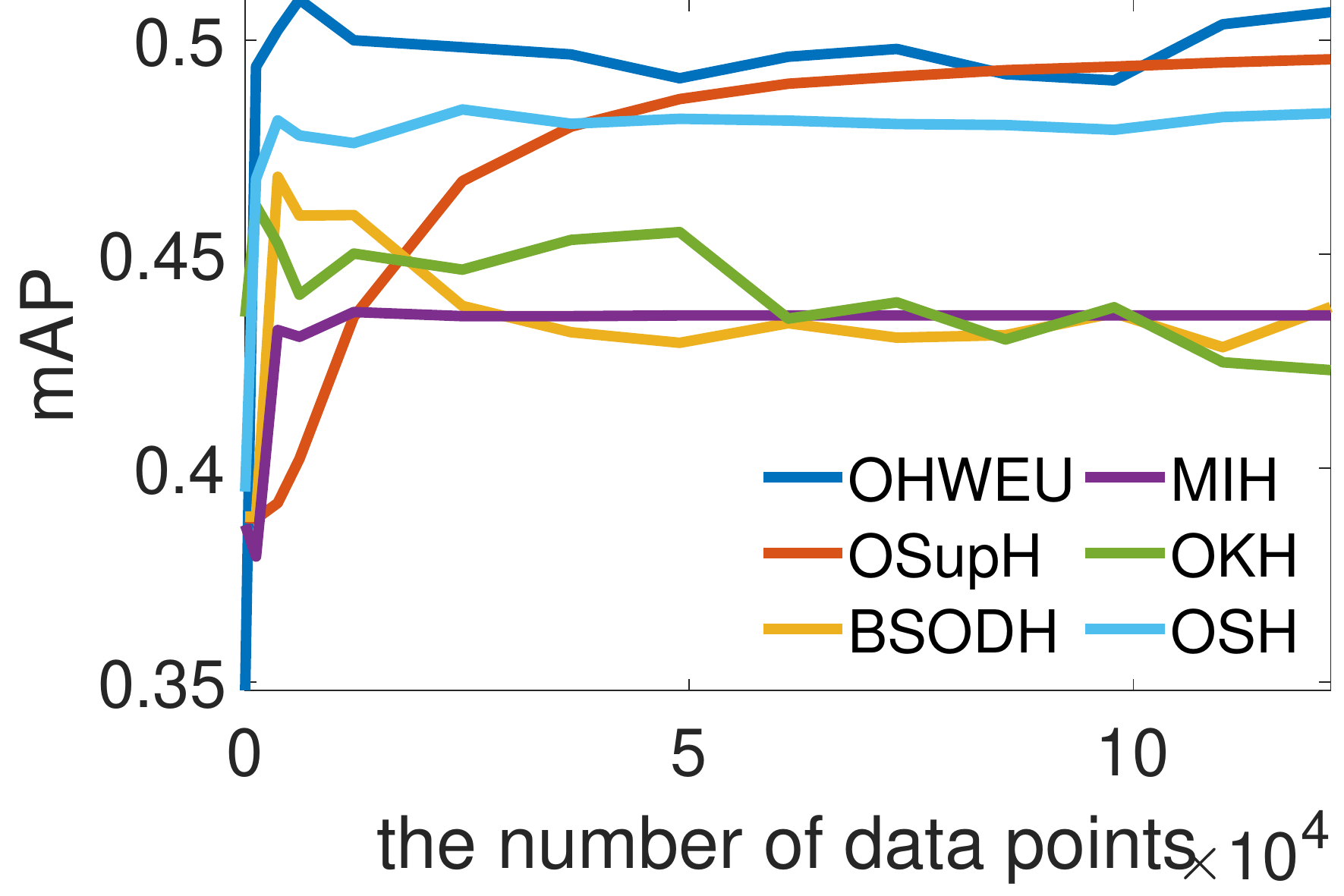}&
 \includegraphics[width=0.48\columnwidth]{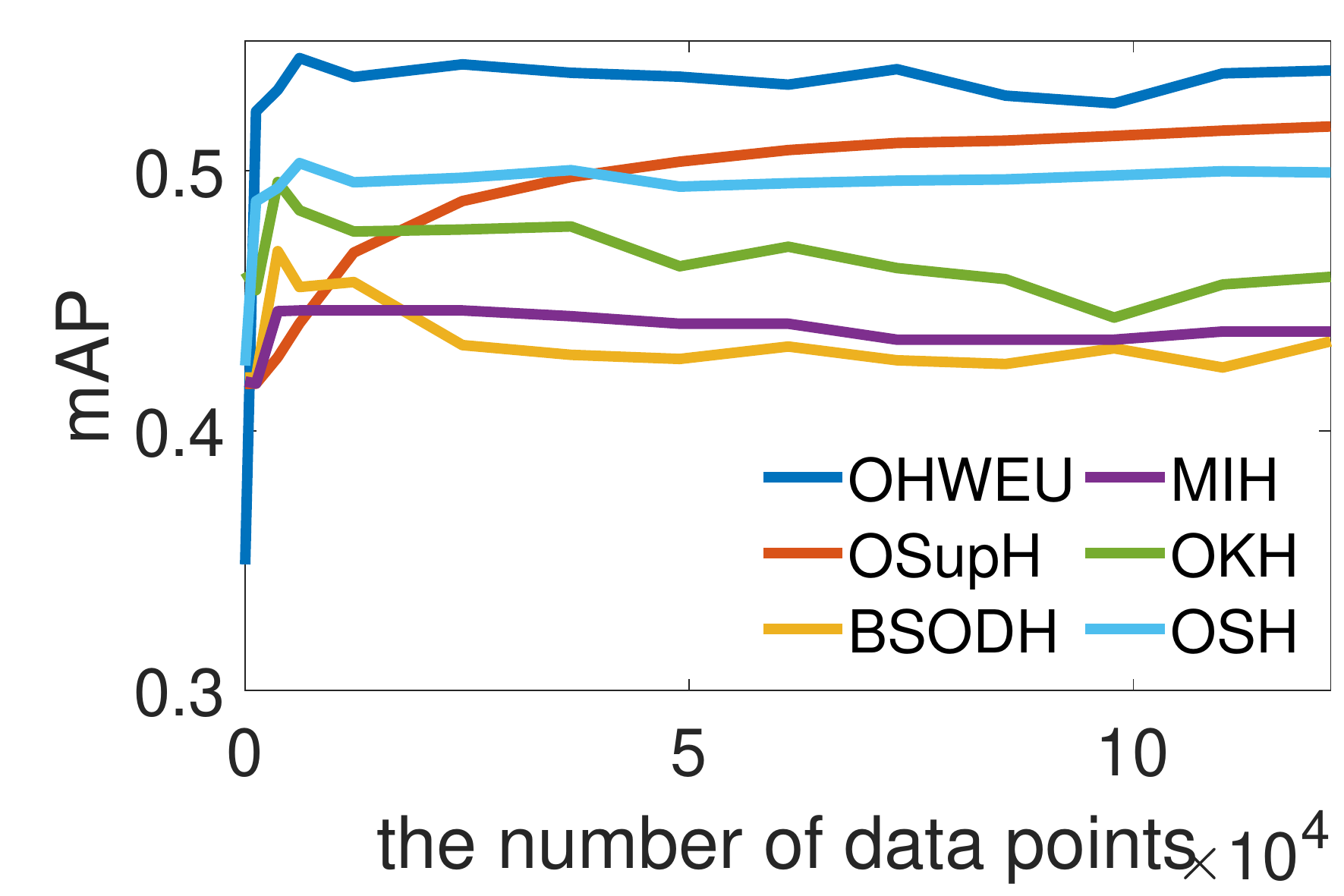}&
 \includegraphics[width=0.48\columnwidth]{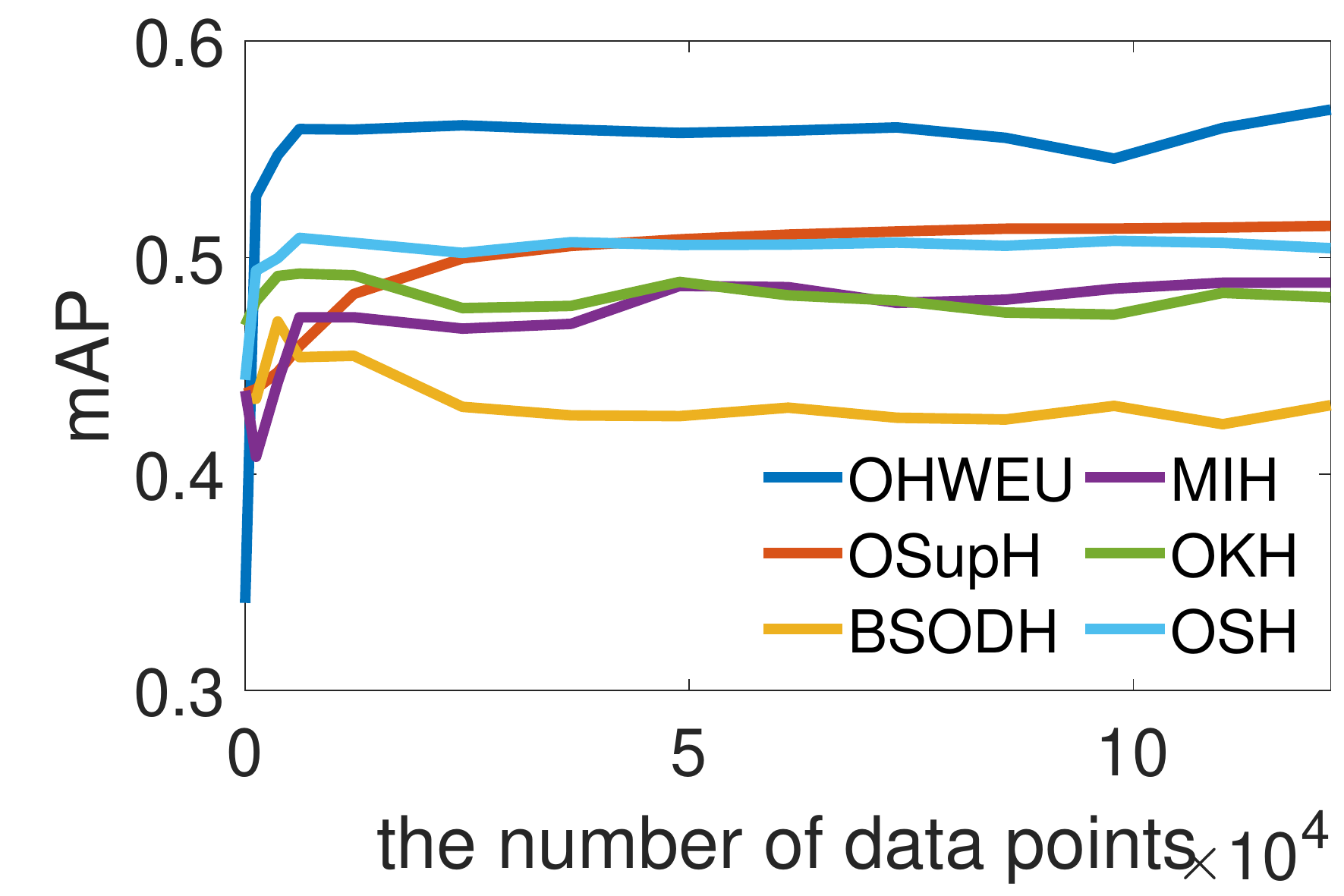}&
 \includegraphics[width=0.48\columnwidth]{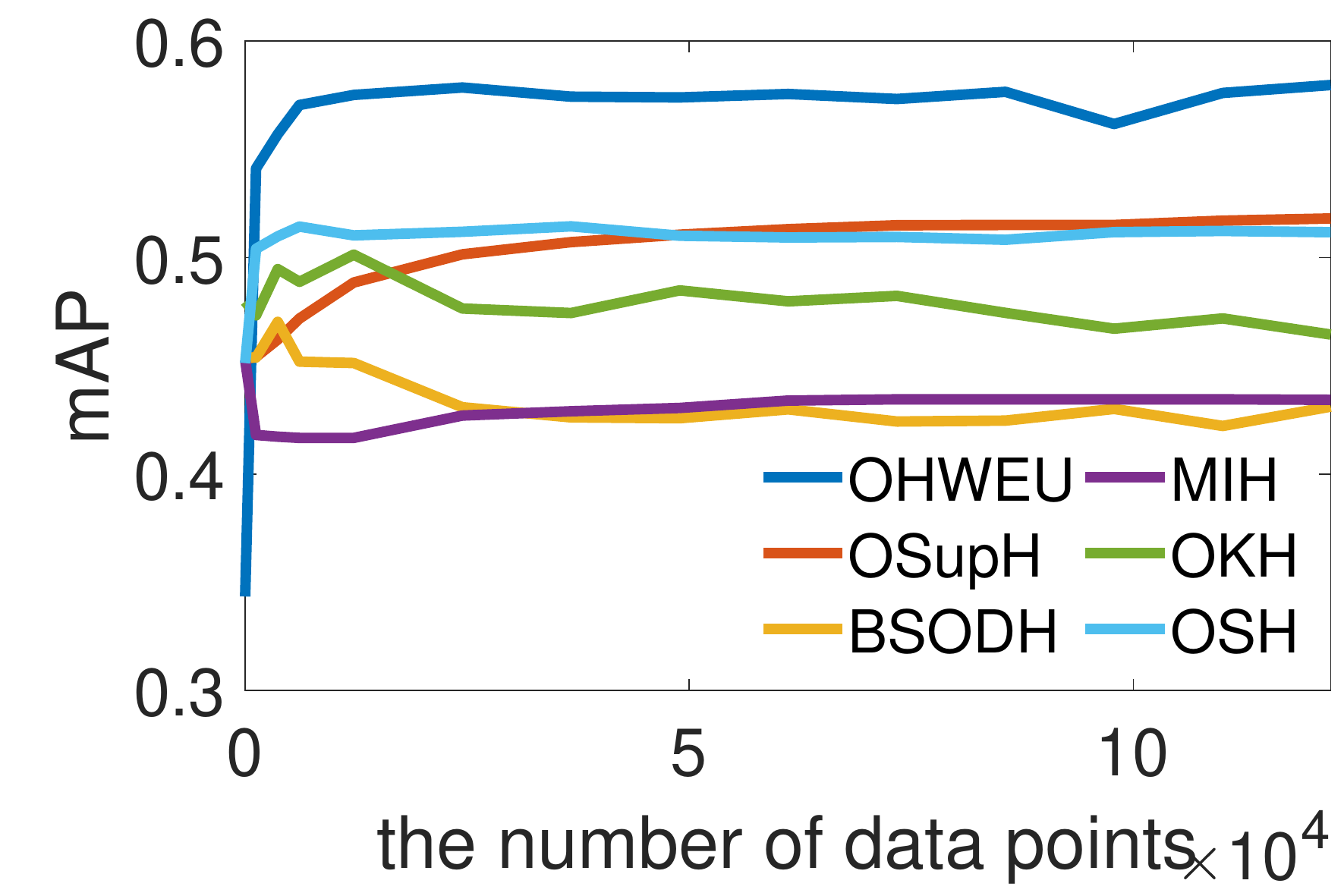}\\
{\small (a) 32 bits}  &  {\small (b) 64 bits} & {\small (c) 96 bits}  &  {\small (d) 128 bits}
\end{tabular}

\caption{The search performance of different methods with training data increasing on MS-COCO.}
\label{fig:res3}

\end{figure*}
\subsection{Comparison of Accuracy}
We compare our method with BSODH~\cite{DBLP:conf/aaai/LinJLSWW19}, MIH~\cite{Cakir2017MIHash}, OSupH~\cite{cakir2017online}, OKH~\cite{7907165}, OSH~\cite{leng2015online}, The key parameters in each compared method are set as the ones recommended in the corresponding papers. In MIH, the reservoir size is set to 200 as the reservoir size is related to the training time cost of MIH. As OKH does, we have $m$ = 300 data points to train the hash functions in the initial stage.

Fig.~\ref{fig:res2} shows the mAP results from 32 bits to 128 bits. On MS-COCO, our method outperforms other online hashing methods. And OSupH is the second best online hashing method. On NUS-WIDE, our method is still the best method among the online hashing methods. From 32 bits to 96 bits, MIH is the second best online hashing methods, while OSupH is the second best online hashing method for 128 bits.

Fig.~\ref{fig:res3} shows the mAP performance of different online hashing methods with the training data increasing from 32 bits to 128 bits on MS-COCO. According to the results, we can see that our method outperforms other online hashing methods after taking only a few training data for training. And our method has a stable generalization ability as our method can achieve the satisfactory performance with only a few training data.

\begin{figure*}[t]
\centering
\begin{tabular}{cccc}
 \includegraphics[width=0.48\columnwidth]{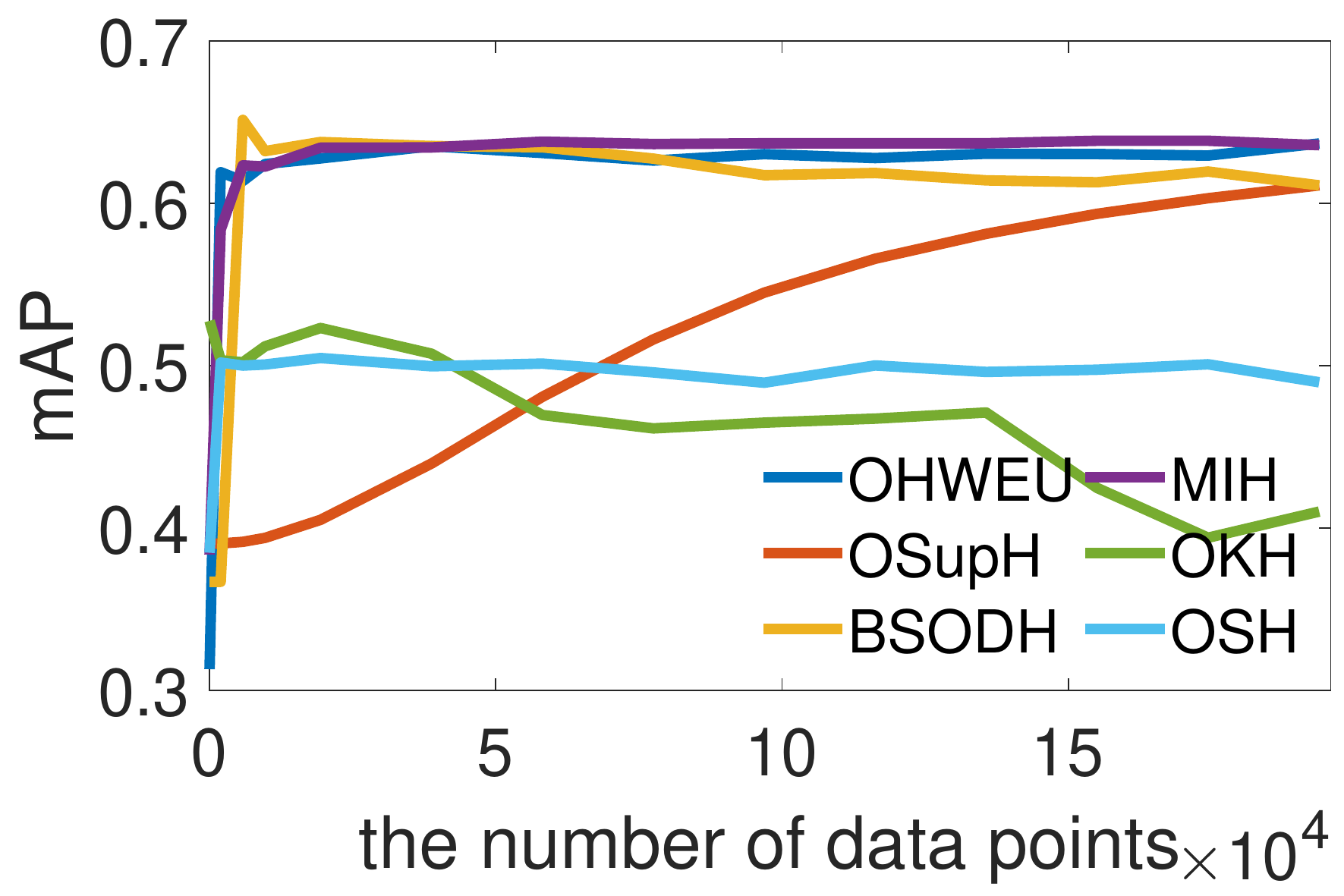}&
 \includegraphics[width=0.48\columnwidth]{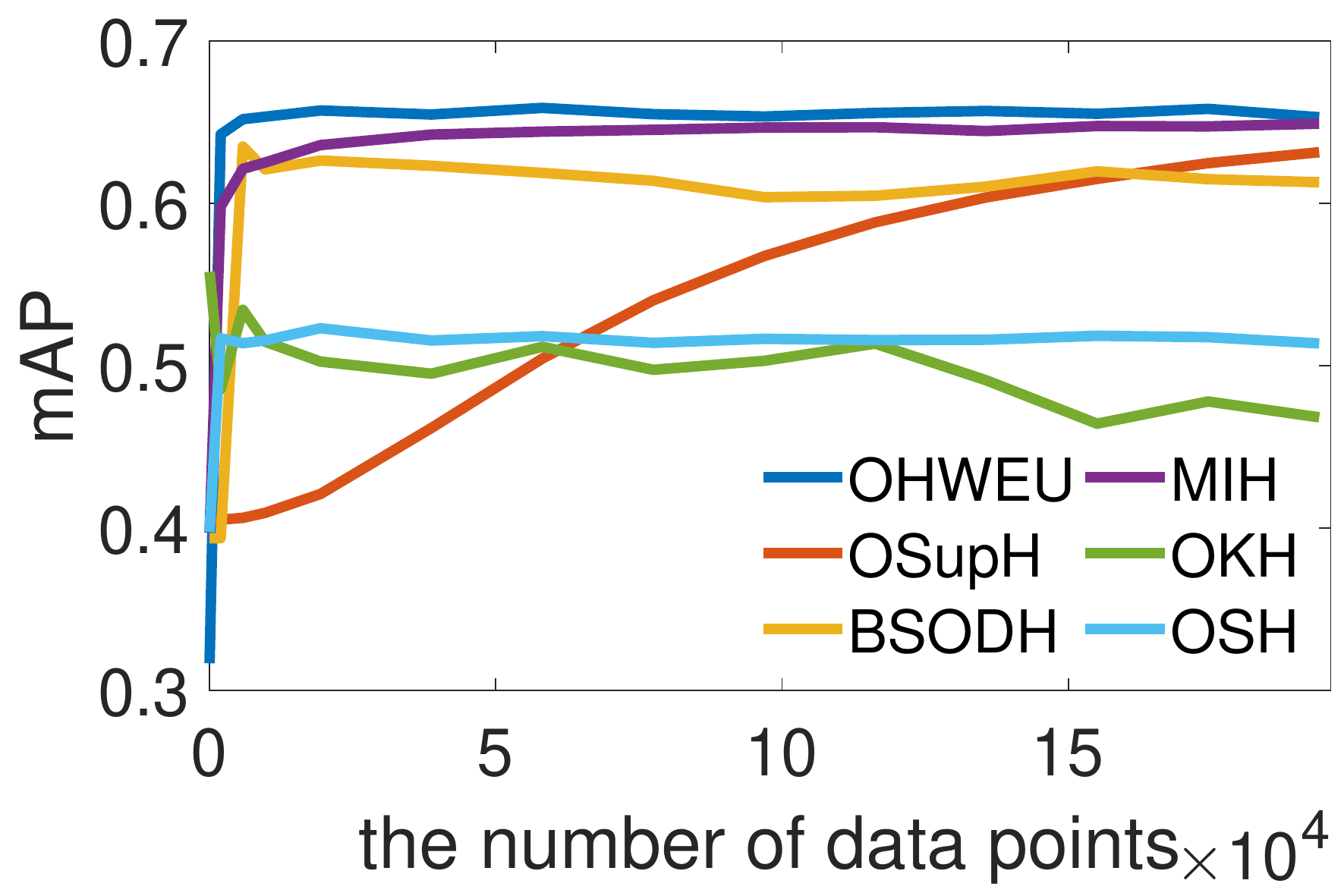}&
 \includegraphics[width=0.48\columnwidth]{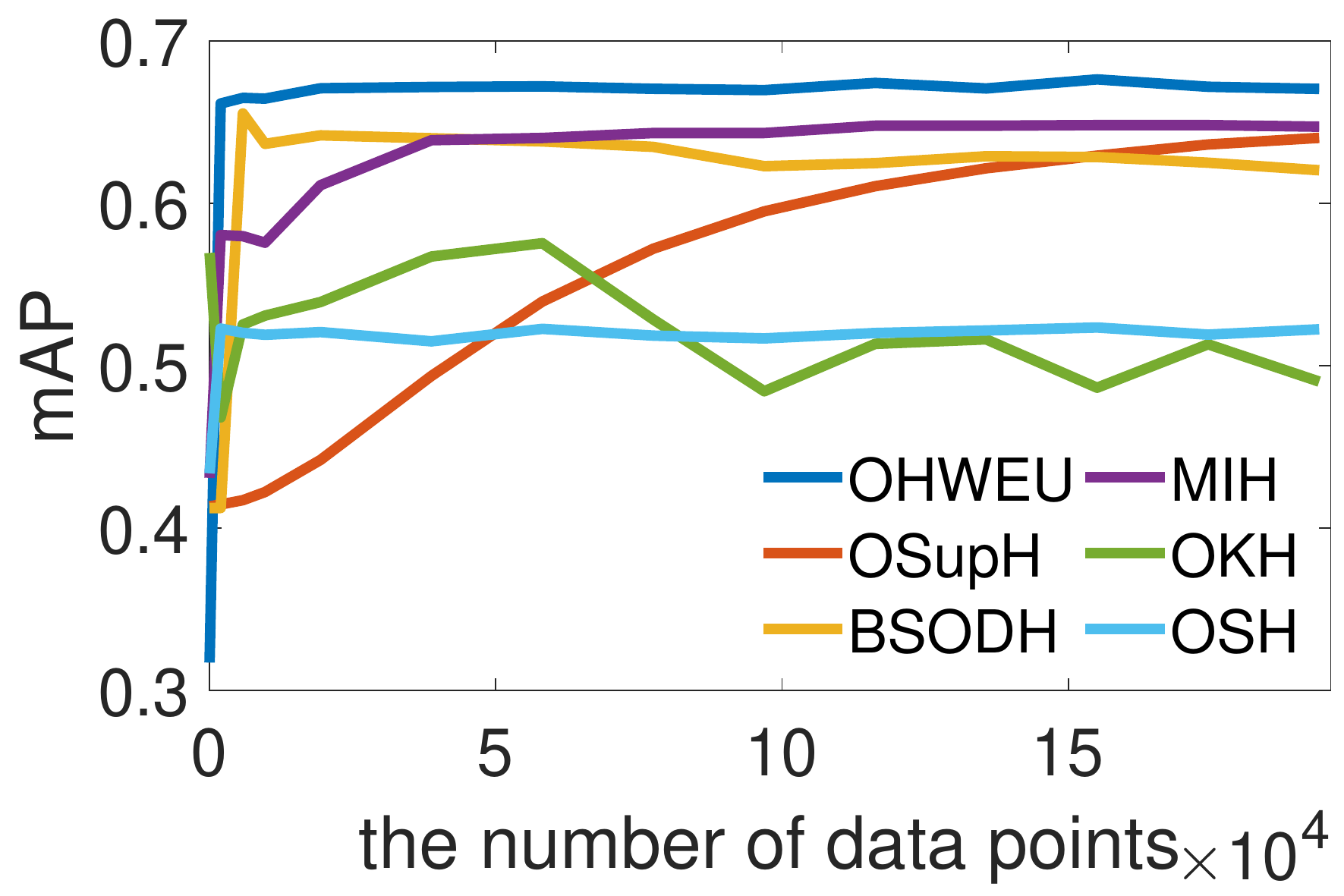}&
 \includegraphics[width=0.48\columnwidth]{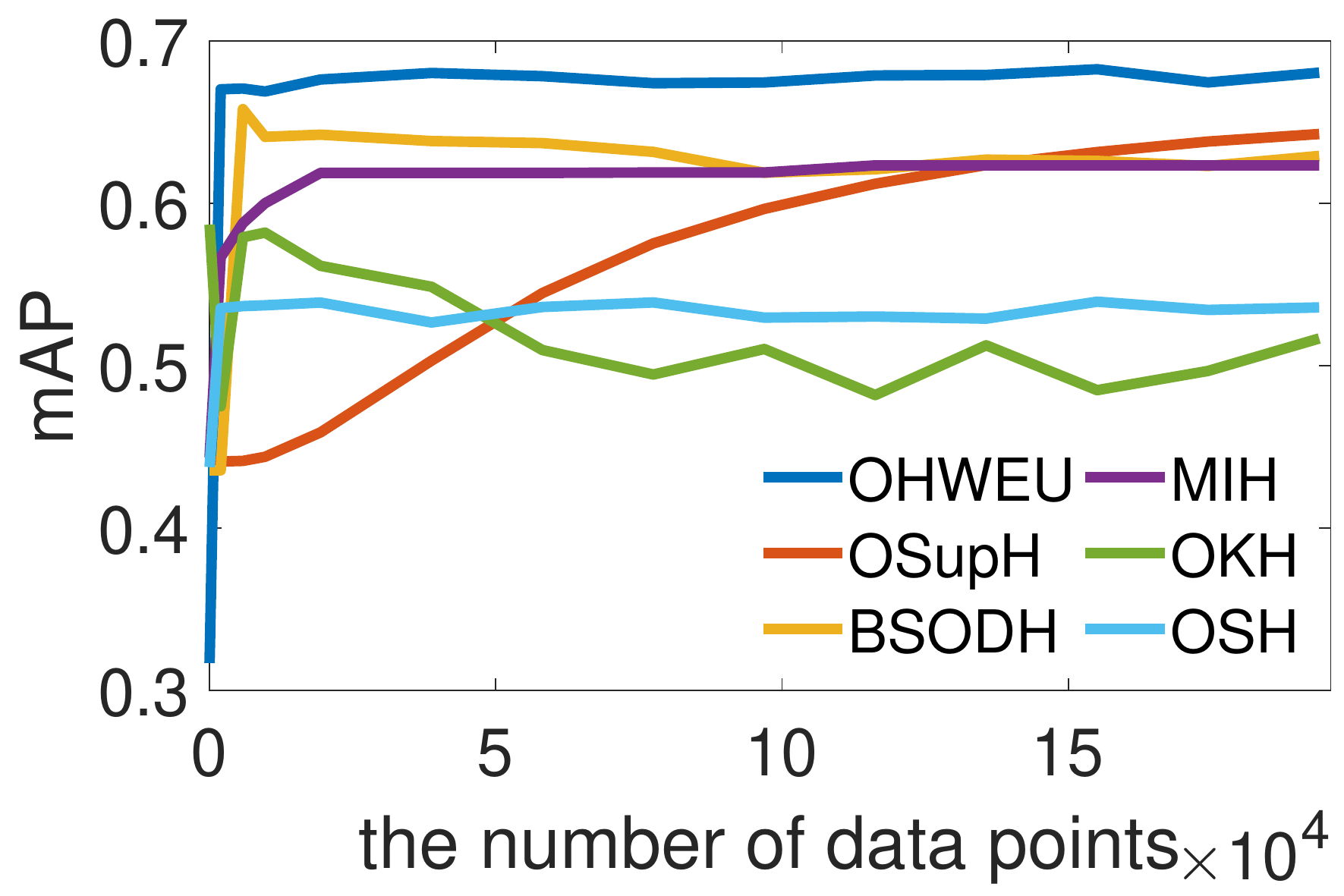}\\
{\small (a) 32 bits}  &  {\small (b) 64 bits} & {\small (c) 96 bits}  &  {\small (d) 128 bits}
\end{tabular}

\caption{The search performance of different methods with training data increasing on NUS-WIDE.}
\label{fig:res4}
\end{figure*}

Fig.~\ref{fig:res4} shows the mAP performance of different online hashing methods with the training data increasing from 32 bits to 128 bits on NUS-WIDE. For 32 bits, the curves of our method, MIH, and BSODH are very close, our method has a close performance to MIH and BSODH. From 64 bits to 128 bits, our method obviously outperforms MIH, BSODH, and other online hashing methods.

\begin{figure*}[t]
\centering
\begin{tabular}{cccc}
 \includegraphics[width=0.48\columnwidth]{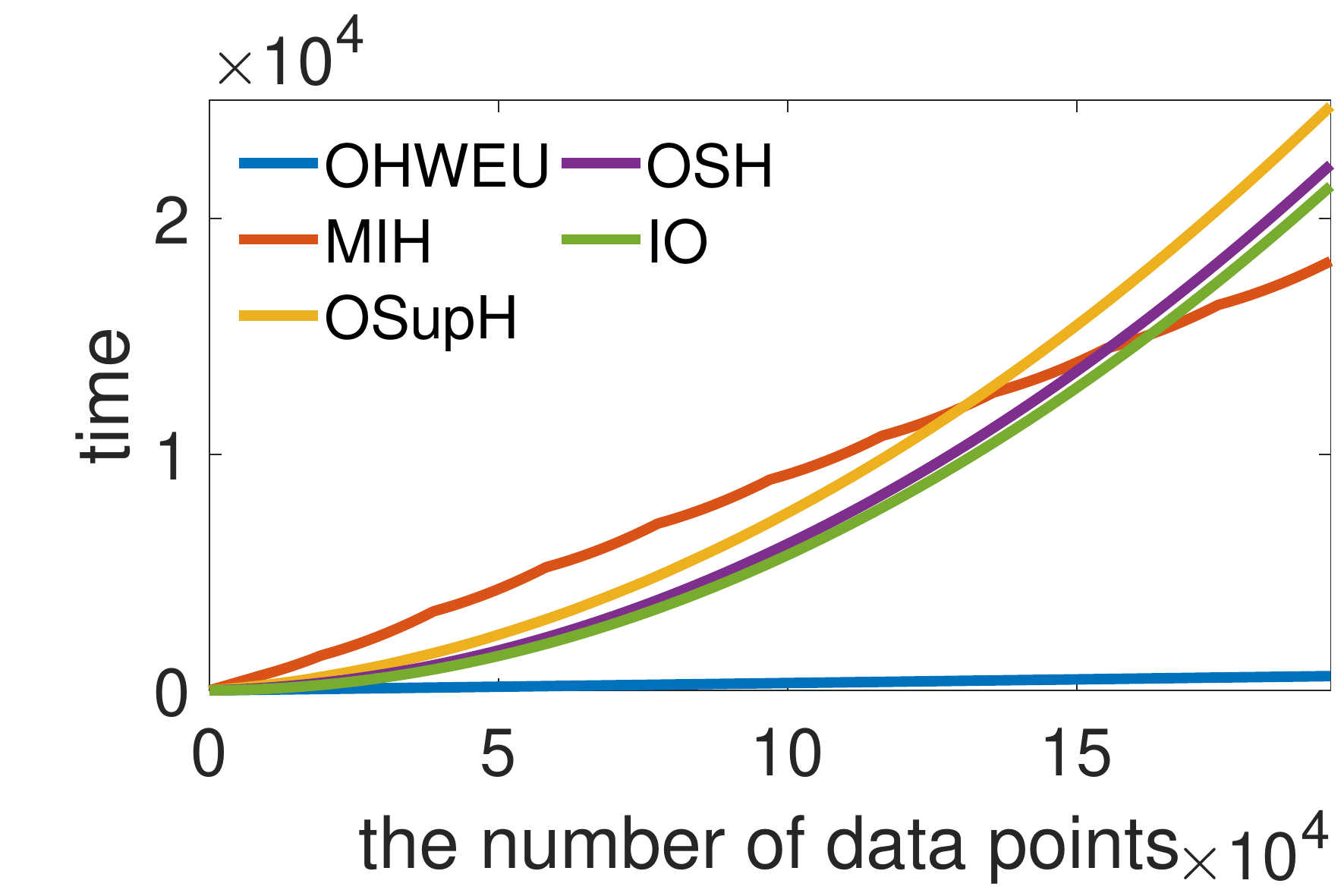}&
 \includegraphics[width=0.48\columnwidth]{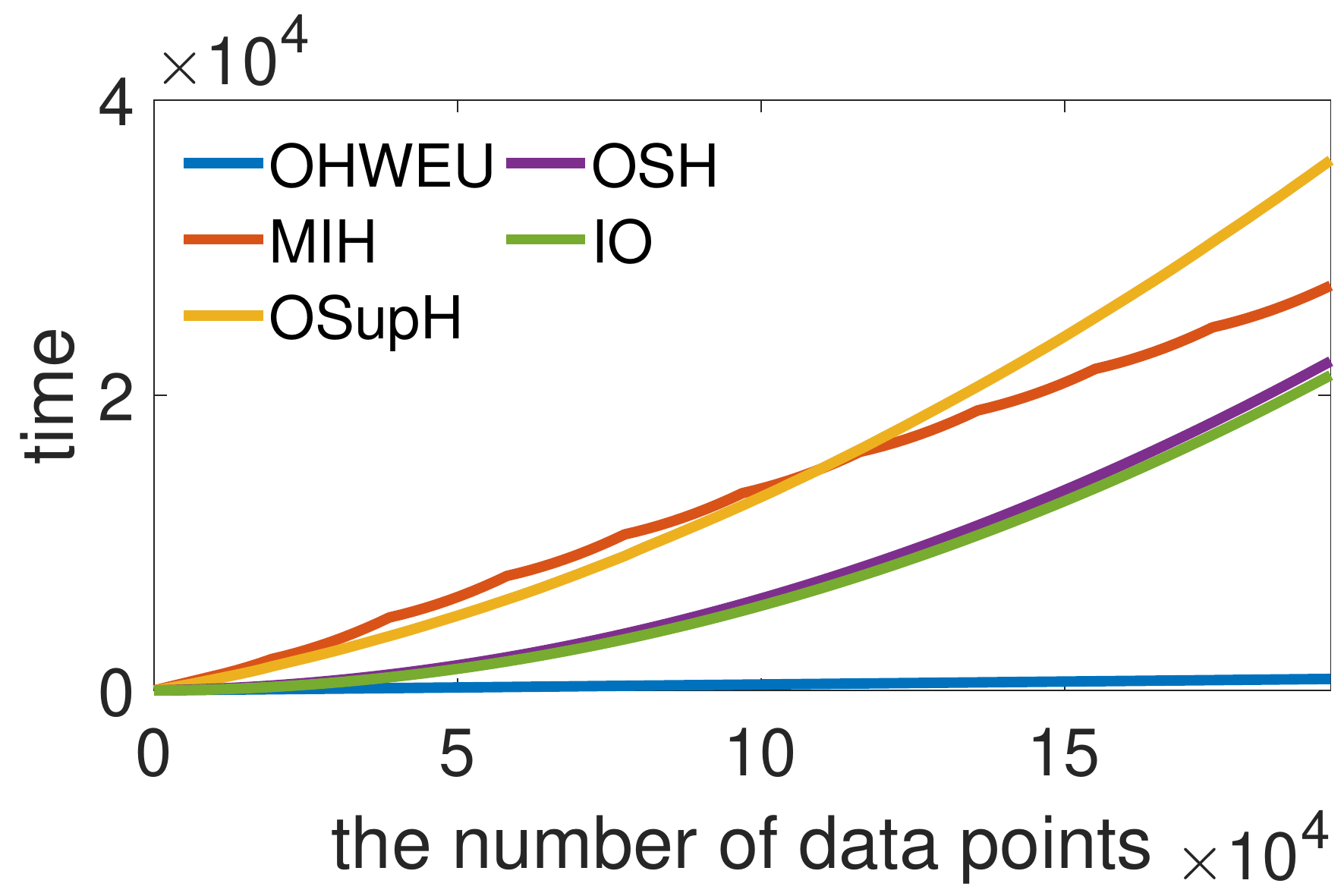}&
 \includegraphics[width=0.48\columnwidth]{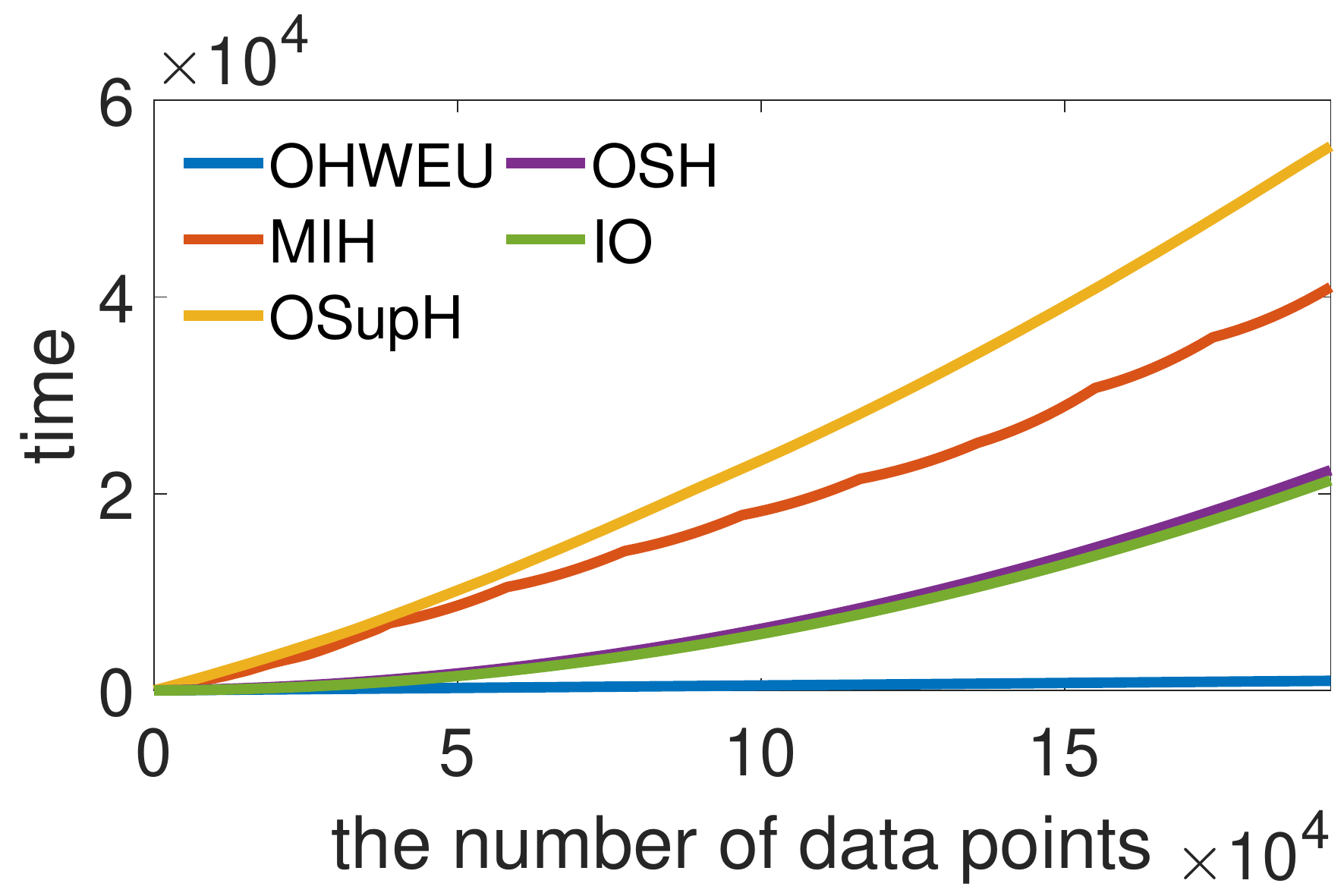}&
 \includegraphics[width=0.48\columnwidth]{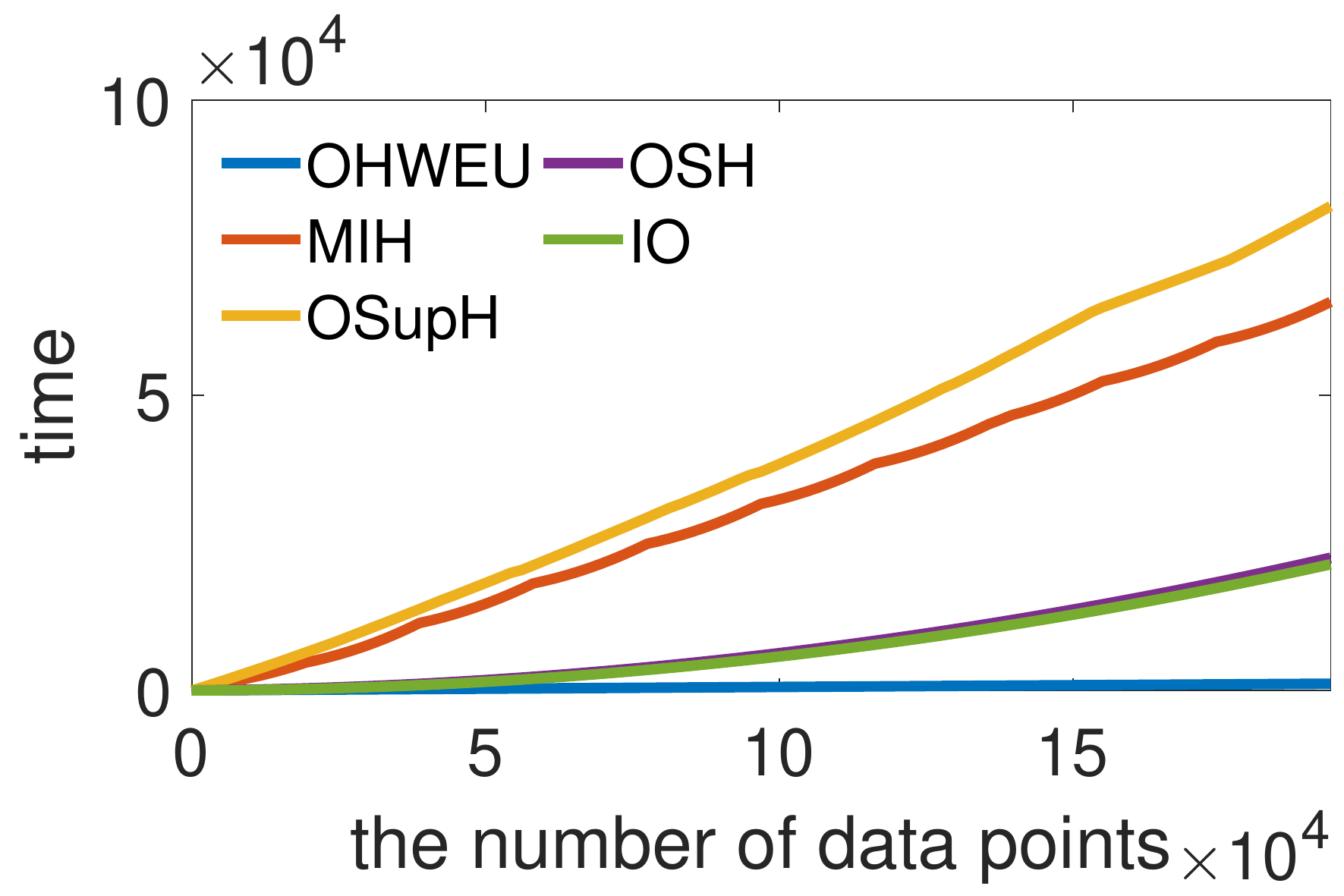}\\
{\small (a) 32 bits}  &  {\small (b) 64 bits} & {\small (c) 96 bits}  &  {\small (d) 128 bits}
\end{tabular}

\caption{The time cost of different methods with training data increasing on NUS-WIDE.}
\label{fig:res5}
\end{figure*}

\subsection{Comparison of Efficiency}
Table~\ref{table:teaser2} and~\ref{table:teaser3} show the training time of different online hashing methods on NUS-WIDE and MS-COCO, respectively. From the results, we can see that our method is faster than other online hashing methods. OSupH and MIH are much slower than other hashing methods. In addition to learning the hash functions, MIH introduces a trigger update module to determine whether to update the binary codes by evaluating the performance of the hash functions on the reservoir sample, and the calculation in the trigger update module is time-consuming. OSupH adopts the online boosting algorithm and needs to consider the hash functions on the previous bits when learning one hash function, which is inefficient, especially for the long bits.

\begin{table}[!ht]
	\caption{The training time (in second) on NUS-WIDE.}
	\begin{center}
	\centering
	\begin{tabular}{m{0.16\columnwidth}<{\centering}m{0.14\columnwidth}<{\centering}m{0.14\columnwidth}<{\centering}m{0.14\columnwidth}<{\centering}m{0.14\columnwidth}<{\centering}}\hline\hline

            & 32 bits & 64 bits & 96 bits & 128 bits  \\ \hline
    \centering{$\bold{OHWEU}$}      &  $\bold{181.34}$  &	$\bold{353.84}$	& $\bold{550.40}$ & $\bold{687.52}$\\ \hline
    \centering{OSH}     & 484.39 & 524.09 & 599.66 &  693.28  \\
    \centering{OSupH}     & 2964.77 & 14146.94  & 33543.83 & 60548.74 \\
    \centering{MIH}     & 17759.40 & 27008.53 & 40120.89 & 64800.35   \\
    \centering{BSODH}     & 2116.42 & 2404.66 & 2881.74 & 3498.40 \\\hline
 		\end{tabular}
	\end{center}

	\label{table:teaser2}
\end{table}

\begin{table}[!ht]
	\caption{The training time (in second) on MS-COCO.}
	\begin{center}
	\centering
    \begin{tabular}{m{0.16\columnwidth}<{\centering}m{0.14\columnwidth}<{\centering}m{0.14\columnwidth}<{\centering}m{0.14\columnwidth}<{\centering}m{0.14\columnwidth}<{\centering}}\hline\hline

            & 32 bits & 64 bits & 96 bits & 128 bits  \\ \hline
    \centering{$\bold{OHWEU}$}      &  $\bold{125.12}$  &	$\bold{245.35}$	& $\bold{208.02}$ & $\bold{466.08}$\\ \hline
    \centering{OSH}     & 335.28 & 387.80 & 485.61 & 606.08 \\
    \centering{OSupH}     & 3650.17 & 12230.75 & 25880.18 & 53348.59  \\
    \centering{MIH}     & 43091.97 & 60450.99 & 73379.56 & 80787.40    \\
    \centering{BSODH}     & 3052.27 & 3251.55 & 3617.38 & 3786.90  \\\hline\hline
 		\end{tabular}
	\end{center}

	\label{table:teaser3}
\end{table}

Following the setting in ~\cite{leng2015online}, we assume the data come in chunks. Each chunk is composed of 1000 data points. We simulate a situation that exchanges the data between the hard disk and RAM to compute the new binary codes, which is implemented by C++ code. It takes 3.97s to transfer 1000 4096-D data points from the hard disk to RAM averagely. Fig.~\ref{fig:res5} shows the accumulated time cost of different online hashing methods with the data samples increasing for different online hashing methods on NUS-WIDE from 32 bits to 128 bits. 'IO' denotes the time cost of accumulating all the received data to update the binary codes. The experiments are run on a PC with Intel i7 3.4 GHz CPU, 24 GB memory.

According to the results in Fig.~\ref{fig:res5}, our method has the lowest time cost among the online hashing methods, since our method has a low time cost for learning the projection functions and does not need to accumulate the whole database to update the binary codes. For OSupH, OSH, and IO, the accumulated time cost rises as an quadratic function of the number of data. Comparing OSH with IO, we can see that most of the time cost of OSH comes from transferring the data from the disk to RAM for the updating of the binary codes. MIH adopts the trigger update module to automatically determine when to recompute the binary codes, and its time cost is approximately linear to the data size. However, its time cost is still high as using the trigger update module to determine whether to update is time-consuming.

\subsection{Parameter Analysis}
\begin{figure}[t]
\centering
\begin{tabular}{cc}
 \includegraphics[width=0.455\columnwidth]{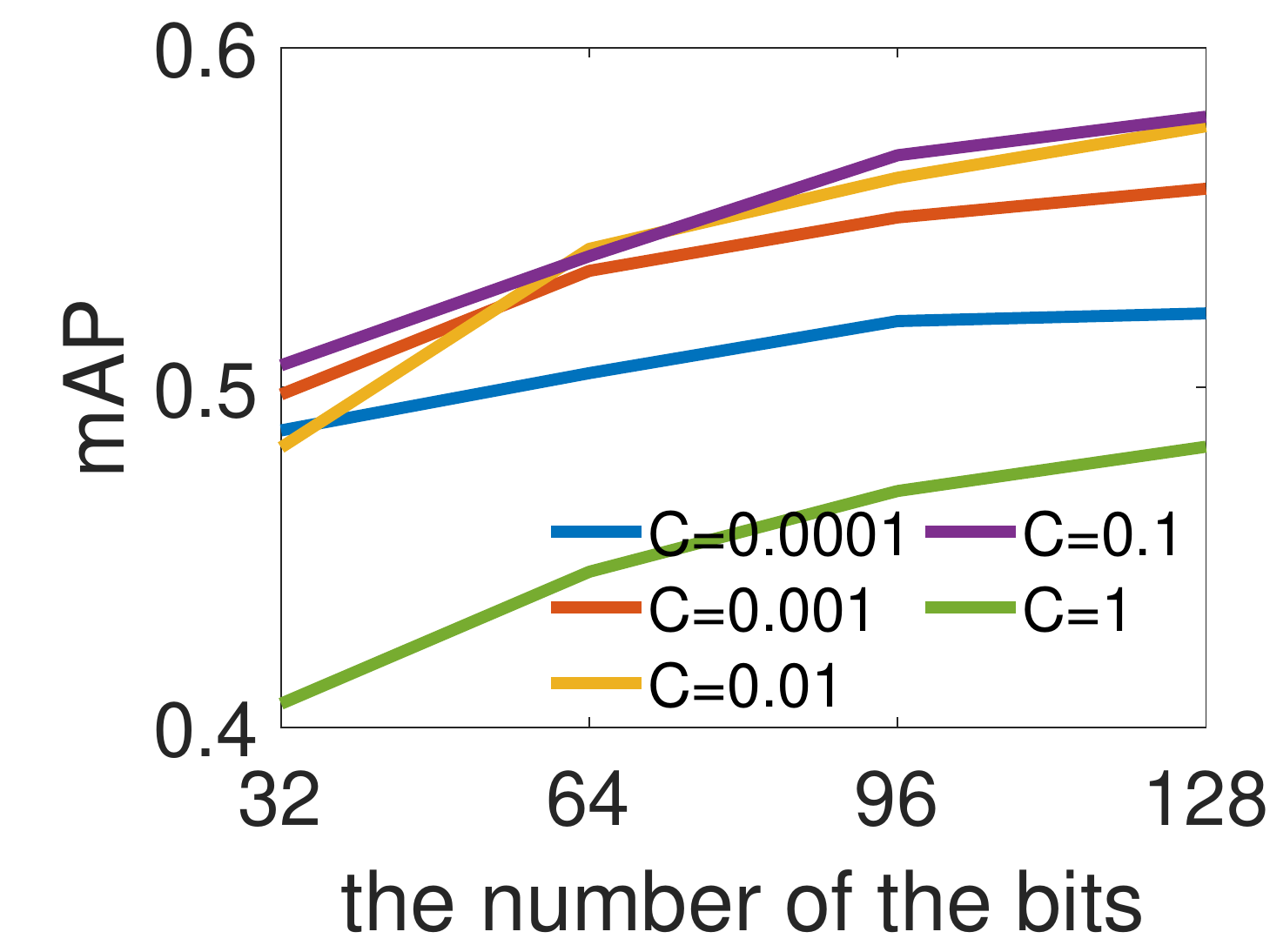}&
 \includegraphics[width=0.455\columnwidth]{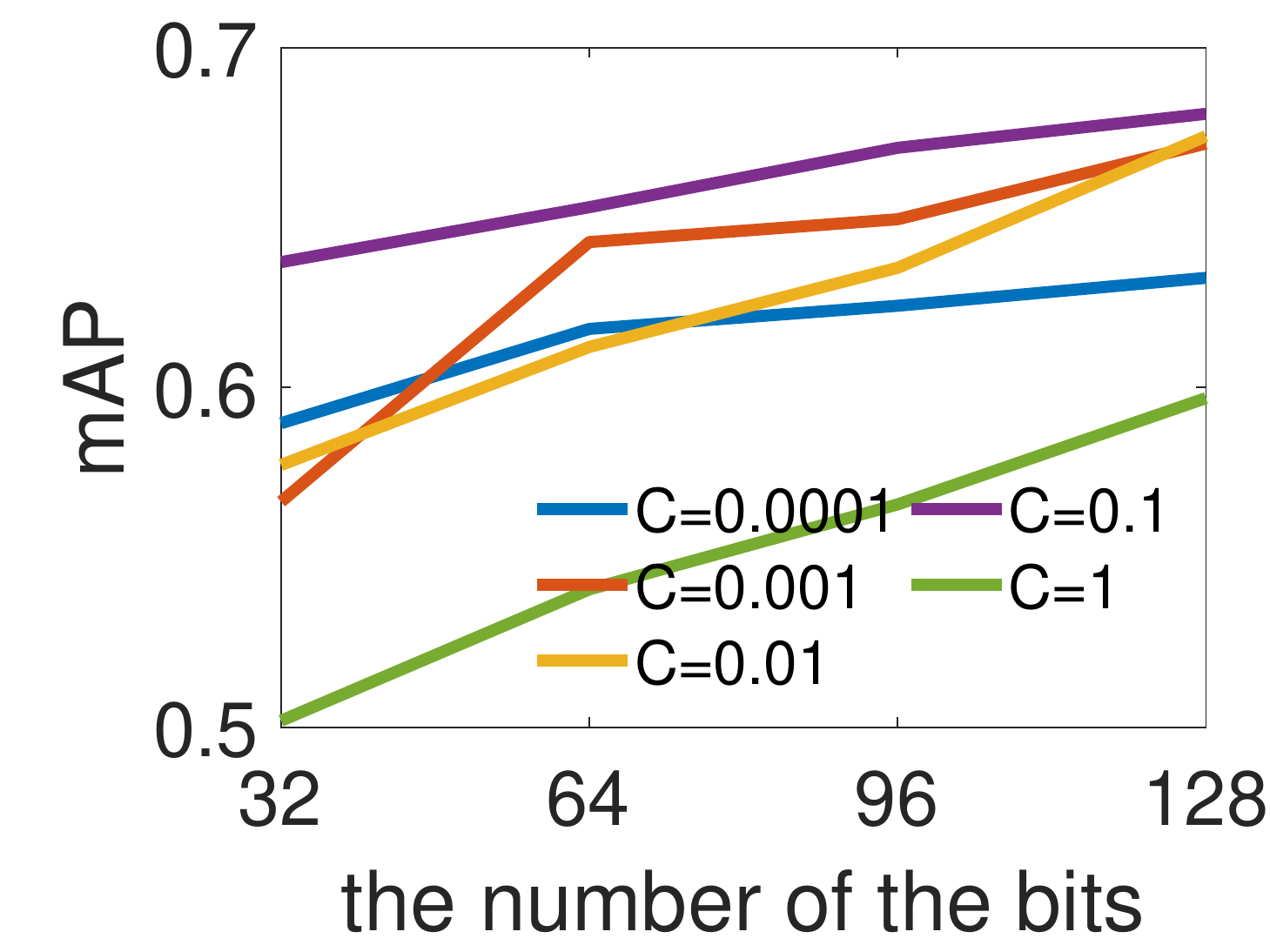}\\
{\small (a) MS-COCO}  &  {\small (b) NUS-WIDE}
\end{tabular}

\caption{The mAP results with different $C$.}
\label{fig:res6}

\end{figure}

As indicated in Eqn. (6), $C$ is used to control the trade-off between maintaining the projection vectors close to the previous ones and minimizing the current loss. We investigate the influence of the parameter $C$. Fig.~\ref{fig:res6} shows our method with different $C$ on MS-COCO and NUS-WIDE. From the results, we can see that on both datasets, the performance of the proposed method becomes better when $C$ increases at first. However, when $C$ reaches some threshold, further increasing $C$ leads to the performance degradation. Based on the results, we set $C = 0.1$ for MS-COCO and NUS-WIDE.

\begin{figure}[t]
\centering
\begin{tabular}{cc}
 \includegraphics[width=0.455\columnwidth]{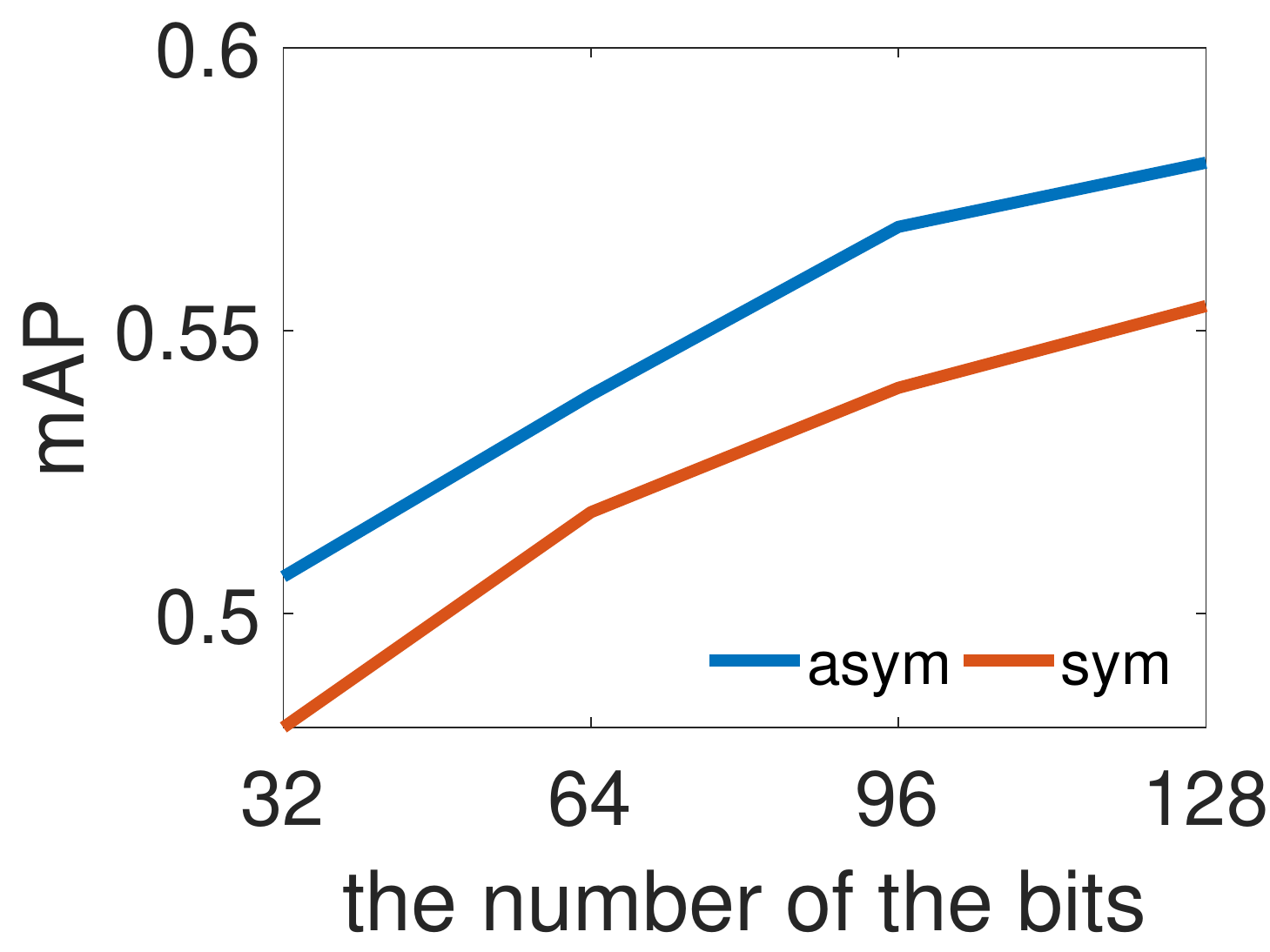}&
 \includegraphics[width=0.455\columnwidth]{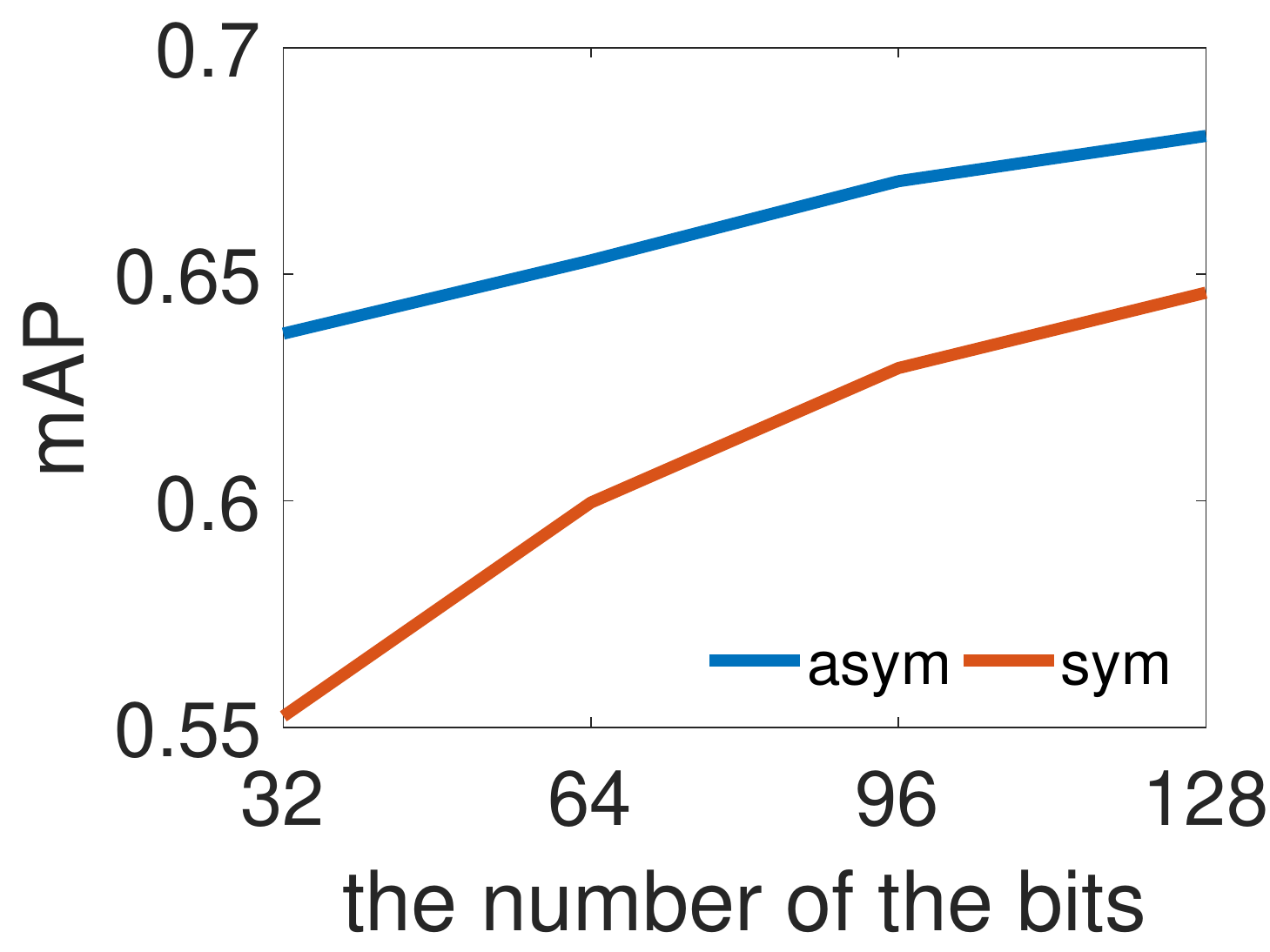}\\
{\small (a) MS-COCO}  &  {\small (b) NUS-WIDE}
\end{tabular}

\caption{The comparison between symmetric processing and asymmetric processing.}
\label{fig:res7}

\end{figure}

Fig.~\ref{fig:res7} shows the comparison of our method with different ways to process the queries and the binary code database. ``sym'' denotes that our method treats the queries and the binary code database symmetrically, and compares the query with the database points according to Eqn. (15). ``asym'' denotes that our method uses an asymmetric way and compares the query and the database points according to Eqn. (25). According to the results, ``asym'' outperforms ``sym'', which shows that treating the query and the database asymmetrically can improve the search accuracy.

\section{Conclusions}
In this paper, we propose a new online hashing framework to update the binary codes efficiently without accumulating the whole database. Different from the widely-used online hashing methods, in our framework, the hash functions are fixed and the projection functions are updated from the streaming data to project the binary codes into another binary space. Hence, the binary codes can be updated efficiently. The experiments show that our method can achieve better retrieval accuracy with lower time cost compared to other online hashing methods for multi-label image retrieval.

\section{Acknowledgements}
This work is supported by Shenzhen Municipal Development and Reform Commission (Disciplinary Development Program for Data Science and Intelligent Computing), Key-Area Research and Development Program of Guangdong Province ($\#$2019B010137001), NSFC-Shenzhen Robot Jointed Founding (U1613215).
\bibliographystyle{aaai}
\bibliography{egbib}

\clearpage
\appendixpage
\section{Appendix A. Derivation of $\bold{R}$}
The projection matrix $\bold{R}\in {\mathbb R}^{D \times K}$ is composed of projection vectors $[\bold{r}_1,...,\bold{r}_{K}]$ and each vector corresponds to one bit of the ideal binary code. For notational simplicity, we use $\bold{r}$ to denote the $k^{th}$ vector $\bold{r}_k$. The loss function of $\bold{r}$ for each coming data point $\bold{x}_i$ is defined as:
\begin{equation}
\hat l({g_i^*},{{\bf{x}}_i}) = \left\{ \begin{array}{l}
0\quad \quad \quad \quad \quad {g_i^*}({{\bf{r}}^T}{{\bf{x}}_i}) \ge 1,\\
1 - {g_i^*}({{\bf{r}}^T}{{\bf{x}}_i})\quad otherwise.
\end{array} \right.
\end{equation}

For brevity, we use $\hat l_i$ to denote the loss $\hat l({g_i^*},{{\bf{x}}_i})$ for the $i^{th}$ data point. Then, for each round $i$, we solve the following convex problem with soft margin:
\begin{equation}
\begin{array}{l}
{{\bf{r}}_i} = \arg {\kern 1pt} \;\min \;{\textstyle{1 \over 2}}||{\bf{r}} - {{\bf{r}}_{i - 1}}||^2 + C\xi \\
\;\;\;\;\;s.t.\quad {\hat l_i} \le \xi \;and\;\xi  \ge 0.
\end{array}
\end{equation}

When $\hat l_i$ = 0, $\bold{r}_{i}$ = $\bold{r}_{i-1}$ satisfies Eqn.(2) directly. Otherwise, we define the Lagrangian as:
\begin{equation}
\begin{split}
L(\bf{r},\tau,\lambda,\xi) &= {\textstyle{1 \over 2}}||{\bf{r}} - {{\bf{r}}_{i - 1}}||^2 + C\xi \\
                           &+ \tau (1-\xi-{g_i^*}({{\bf{r}}^T}{{\bf{x}}_i})) - \lambda\xi,\\
\end{split}
\end{equation}
with $\tau \ge 0$ and  $\lambda \ge 0$ are Lagrange multipliers.

Let ${\textstyle{\partial L(\bf{r},\tau,\lambda,\xi)  \over \partial \bold{r}}} = 0$. We have
\begin{equation}
{\textstyle{\partial L(\bf{r},\tau,\lambda,\xi)  \over \partial \bold{r}}} = \bold{r} - \bold{r}_{i-1} - \tau{g^*_i}{\bf{x}}_i = 0.
\end{equation}

Let ${\textstyle{\partial L(\bf{r},\tau,\lambda,\xi)  \over \partial \xi}} = 0$. We have
\begin{equation}
{\textstyle{\partial L(\bf{r},\tau,\lambda,\xi)  \over \partial \xi}} = C - \tau - \lambda = 0.
\end{equation}

As $\lambda \ge 0$, $\tau \le C$.

Plugging Eqn.(4) and (5) back in Eqn.(3), we obtain
\begin{equation}
L(\tau)= -{\textstyle{1 \over 2}}\tau^2(g^*_i)^2||\bold{x}_i||^2+\tau(1-{g_i^*}({{\bf{r}}_{i-1}^T}{{\bf{x}}_i})).
\end{equation}

As $g^*_i \in \{1, -1\}$, $(g^*_i)^2=1$. Let ${\textstyle{\partial L(\tau)  \over \partial \tau}} = 0$. We have
\begin{equation}
\tau = {\textstyle{ 1-{g_i^*}({{\bf{r}}_{i-1}^T}{{\bf{x}}_i})  \over ||\bold{x}_i||^2}}.
\end{equation}

Hence, the optimal $\bold{r}$ is
\begin{equation}
{{\bf{r}}_i} = {{\bf{r}}_{i - 1}} + \tau {g^*_i}{{\bf{x}}_i},
\end{equation}
where
\begin{equation}
\tau  = \min \{ C,\frac{{1-{g_i^*}({{\bf{r}}_{i-1}^T}{{\bf{x}}_i})}}{{||\bold{x}_i||^2}}\}.
\end{equation}

\section{Appendix B. Proof of $\bold{Theorem}$ $\bold{2}$}
$\bold{Theorem}$ $\bold{2}$ Let ($\bold{x}_1$, $g_1^*$),$\cdots$,($\bold{x}_i$, $g_i^*$) be a sequence of pair-wise examples with label $g_i^* \in \{-1,1\}$ for all $i$. As before, Eqn.(26) in the manuscript denotes by $\hat l_i=\hat l(\bold{r}; (\bold{x}_i$, $g_i^*))$ the instantaneous loss suffered by our algorithm on round $i$. Let $\bold{u}$ be an arbitrary vector that $\bold{u} \in {\mathbb R}^{K}$, and define $\hat l_i^*=\hat l(\bold{u}; (\bold{x}_i$, $g_i^*))$ as the loss suffered by $\bold{u}$. Then, by assuming $||{\bf{x}}_i|| \le R$, for any vector $\bold{u} \in {\mathbb R}^K$, the number of prediction mistakes made by our method on this sequence of examples is bounded by
\begin{equation}
\max \{ {R^2},1/C\} (||\bold{u}|{|^2} + 2C\sum\limits_{i = 1}^t {\hat l_i^*} ).
\end{equation}

$\bold{Proof}$ As $\tau  = \min \{ C,\frac{{1-{g_i^*}({{\bf{r}}_{i-1}^T}{{\bf{x}}_i})}}{{||\bold{x}_i||^2}}\}$, by assuming $||{\bf{x}}_i|| \le R$, we have $\tau  = \min \{ C,\frac{{1-{g_i^*}({{\bf{r}}_{i-1}^T}{{\bf{x}}_i})}}{{R^2}}\}$

According to the definition of the loss function, if our method makes a prediction mistake on round $i$ then $\hat l_i \ge 1$. Therefore,  we have
\begin{equation}
\min \{ 1/{R^2},{\rm{C}}\}  \le {\tau _i}{\hat l_i}.
\end{equation}

Let $M$ be the number of prediction mistakes made on the entire sequence. We have,
\begin{equation}
\min \{ 1/{R^2},{\rm{C}}\} M \le \sum\limits_{i = 1}^t {{\tau _i}{\hat l_i}}.
\end{equation}

Since $\tau  = \min \{ C,\frac{{1-{g_i^*}({{\bf{r}}_{i-1}^T}{{\bf{x}}_i})}}{{R^2}}\}$, we know that
\begin{equation}
\tau_i||\bold{x}_i|{|^2} \le {\hat l_i},
\end{equation}
and
\begin{equation}
\tau_i\hat l_i^* \le C\hat l_i^*.
\end{equation}

According to $\bold{Lemma}$ 1 in ~\cite{crammer2006online}, we have
\begin{equation}
\sum\limits_{i = 1}^t {{\tau _i}(2{\hat l_i} - {\tau _i}||{\bold{x}_i}|{|^2} - 2\hat l_i^*) \le ||\bold{u}|{|^2}}.
\end{equation}

Plugging Eqn.(13) and (14) into Eqn.(15), we obtain
\begin{equation}
\sum\limits_{{\rm{i}} = 1}^t {{\tau _i}{\hat l_i}}  \le ||\bold{u}|{|^2} + 2C\sum\limits_{i = 1}^t {\hat l_i^*}.
\end{equation}

Combining Eqn.(12) with Eqn.(16), we obtain

\begin{equation}
\min \{ 1/{R^2},C\} M \le ||\bold{u}|{|^2} + 2C\sum\limits_{i = 1}^t {\hat l_i^*}.
\end{equation}

Multiplying both sides of the above by $\max \{ {{R^2}},1/{{C}}\}$, we have
\begin{equation}
M \le \max \{ {{R^2}},1/{{C}}\} (||\bold{u}|{|^2} + 2C\sum\limits_{i = 1}^t {\hat l_i^*} ).
\end{equation}

\end{document}